\documentclass
[aps,preprint,10pt,twocolumn,balancelastpage,nofootinbib,superscriptaddress]{revtex4}%
\usepackage{amsfonts}
\usepackage{amsmath}
\usepackage{amssymb}
\usepackage{graphicx}%
\newtheorem{theorem}{Theorem}

\newtheorem{remark}[theorem]{Remark}

\begin{document}
\title{Generalized Fluctuation Theory based on \\the reparametrization invariance of the microcanonical ensemble}
\author{L. Velazquez}
\affiliation{Departamento de F\'{\i}sica, Universidad de Pinar del Rio, Marti 270, Esq. 27
de Noviembre, Pinar del Rio, Cuba}
\pacs{05.70.-a; 05.20.Gg}

\begin{abstract}
The main interest of the present work is the generalization of the
Boltzmann-Gibbs distributions\ and the fluctuation theory based on the
consideration of the reparametrization invariance of the microcanonical
ensemble. This approach allows a novel interpretation of some anomalous
phenomena observed in the non extensive systems like the existence of the
negative specific heats as well as possibilities the enhancing of some Monte
Carlo methods based on the Statistical Mechanics.

\end{abstract}
\date{\today}
\maketitle

\section{Introduction}

We have been witness in the last years of an important reconsideration of the
Foundations of the Thermodynamics and the Statistical Mechanics originated
from the interest to use this kind of description in systems which are outside
the context of their traditional applications. An important area of the
nowadays developments is found in the Thermo-statistical description of the
called \textit{non extensive systems}.

The non extensive systems are those non necessarily composed by a huge number
of constituents, they could be mesoscopic or even small systems, where the
characteristic radio of some underlaying interaction is comparable or larger
than the characteristic linear dimension of the system, particularity leading
to non existence of statistical independence due to the presence of long-range
correlations. In the last years a significant volume of evidences of
thermodynamic anomalous behaviors are been found in the study of the non
screened plasmas and the turbulent diffusion, astrophysical systems, nuclear,
molecular and the atomic clusters, granular matter and complex systems
\cite{bog,sol,lyn,pos,kon,tor,gro1,ato,kud,par,stil}.

It is usual to find in this context equilibrium thermodynamic states
characterized by exhibiting a \textit{negative heat capacity}, that is,
thermodynamic states where an increment of the total energy leads to a
decreasing of the temperature. This behavior is associated to the convexity of
the entropy and the corresponding \textit{ensemble inequivalence} and can
appear as a consequence of the small character of the systems (during the
first-order phase transitions in molecular and atomic or nuclear clusters
\cite{moretto,Dagostino,gro na}) or because of the long-range character of the
interactions (see in the astrophysical systems \cite{antonovb}).

The existence of such anomalous thermodynamical behaviors demands new
developments within the Equilibrium Statistical Mechanics and the
Thermodynamics. An important contribution to the understanding of such
phenomena is found in the formulation of the \textit{Microcanonical
Thermo-statistics }by D. H. E. Gross \cite{gro1}, a theory which returns to
the original basis of the Statistical Mechanics with the reconsideration of
the well-known Boltzmann epitaph:%
\begin{equation}
S=\log W,\label{SB}%
\end{equation}
and allows the study of phase transitions in systems outside the thermodynamic
limit. Since $W$ is the number of microscopic states compatible with a given
macroscopic state, the Boltzmann entropy (\ref{SB}) is a measure of the size
of the microcanonical ensemble. Within Classical Statistical Mechanics $W$ is
just the microcanonical accessible phase space volume, that is, a geometric
quantity. This explains why the entropy (\ref{SB}) does not satisfy the
concavity and the extensivity properties, neither demands the imposition of
the thermodynamic limit or a probabilistic interpretation like the
Shannon-Boltzmann Gibbs extensive entropy:%
\begin{equation}
S_{e}=-\sum_{k}p_{k}\ln p_{k}.\label{sbg}%
\end{equation}
Such geometric character of the Boltzmann entropy leads naturally to the
development of geometric formulations of any statistical formulation derived
from microcanonical basis.

Motived by the successes of the Gross' formulation \cite{gro1}, we have
performed in the refs.\cite{vel.geo,vel-mmc} an analysis about the geometric
features of the microcanonical ensemble. Particularly, we have shown in the
previous paper \cite{vel.geo} that the microcanonical description is
characterized by the presence of an internal symmetry whose existence is
related to the dynamical origin of this ensemble, which has been coined by us
as \textit{the reparametrization invariance}. Such symmetry leads naturally to
the development of a non Riemannian geometric formulation within the
microcanonical description, which leads to an unexpected generalization of the
Gibbs canonical ensemble and the classical fluctuation theory for the open
systems, the improvement of Monte Carlo simulations based on the canonical
ensemble, as well as a reconsideration of any classification scheme of the
phase transitions based on the concavity of the microcanonical entropy.

I shall continuous in the present paper the study of the geometric aspects
related to the existence of the reparametrization invariance. The present work
addresses the study of those equilibrium situations characterized by the
consideration of several control parameters. The main interest will be
focussed in the generalization of the well-known Boltzmann-Gibbs distributions
\ and the fluctuation theory related with the consideration of the
reparametrization invariance ideas. I realize while writing this work that
some of the present ideas have been also rediscovered by Toral in
ref.\cite{toral1} by starting from the Information Theory.

\section{Isolated system}

Let us consider an isolated dynamical system whose macroscopic description
could be carried out by starting from microcanonical basis. Let $\hat{I}%
\equiv\left\{  \hat{I}_{1},\hat{I}_{2}\ldots\mathbf{\ }\hat{I}_{n}\right\}  $
be the set of all those fundamental physical quantities determining the
microcanonical distribution function $\hat{\omega}_{m}$:%
\begin{equation}
\mathbf{\ }\hat{\omega}_{m}\left(  I\right)  =\frac{1}{\Omega\left(  I\right)
}\delta\left[  I-\hat{I}\right]  , \label{mic}%
\end{equation}
which are simultaneously mensurable, that is $\left[  \hat{I}_{i}%
,\mathbf{\ }\hat{I}_{j}\right]  =0$, being $\Omega\left(  I\right)
=Sp\left\{  \delta\left[  I-\hat{I}\right]  \right\}  $ the microcanonical
partition function. Let us suppose that the eigenvalues of the set $\hat{I}$
belong to certain subset $\mathcal{R}_{I}$ of the Euclidean n-dimensional
space $\mathcal{R}^{n}$.

Let us consider another subset $\mathcal{R}_{\Theta}\subset\mathcal{R}^{n}$
which is diffeomorphic equivalent to the subset $\mathcal{R}_{I}$, that is,
there exists a diffeomorphic bijective map $\Theta:\mathcal{R}_{I}%
\rightarrow\mathcal{R}_{\Theta}$ with a non vanishing Jacobian:%
\begin{equation}
\left\vert \frac{\partial\Theta}{\partial I}\right\vert \equiv\det\left[
\frac{\partial\Theta^{k}\left(  I\right)  }{\partial I^{i}}\right]  \not =0,
\end{equation}
for every point $I\in\mathcal{R}_{I}$. It is said that the map $\varphi$
defines a \textit{reparametrization change} of the Euclidean subset
$\mathcal{R}_{I}$. The set of functionals derived from the above map $\Theta$,
$\hat{\Theta}=\left\{  \hat{\Theta}^{1},\hat{\Theta}^{2}\ldots\hat{\Theta}%
^{n}\right\}  =\Theta\left[  \hat{I}\right]  $, determines the same
microscopic Physics of the set $\hat{I}$: every eigenstate $\left\vert
\Psi_{\alpha}\right\rangle $ of $\hat{I}$, $\hat{I}\left\vert \Psi_{\alpha
}\right\rangle =I_{\alpha}\left\vert \Psi_{\alpha}\right\rangle $, is also an
eigenstate of the set $\hat{\Theta}$, $\hat{\Theta}\left\vert \Psi_{\alpha
}\right\rangle =\Theta\left[  \hat{I}\right]  \left\vert \Psi_{\alpha
}\right\rangle =\Theta\left[  I_{\alpha}\right]  \left\vert \Psi_{\alpha
}\right\rangle =\Theta_{\alpha}\left\vert \Psi_{\alpha}\right\rangle $, and
the physical quantities $\hat{\Theta}^{k}$ are also simultaneously mensurable,
$\left[  \hat{\Theta}^{k},\hat{\Theta}^{m}\right]  =0$. There is nothing
strange that the set $\hat{\Theta}$ determines also the \textit{same
macroscopic Physics} of the set $\hat{I}$. Taking into account the well-known
property of the Dirac delta function:%
\begin{equation}
\delta\left[  \Theta-\hat{\Theta}\right]  =\left\vert \frac{\partial\Theta
}{\partial I}\right\vert ^{-1}\delta\left[  I-\hat{I}\right]  \Rightarrow
\Omega\left(  \Theta\right)  =\left\vert \frac{\partial\Theta}{\partial
I}\right\vert ^{-1}\Omega\left(  I\right)  ,
\end{equation}
and therefore:%
\begin{equation}
\frac{1}{\Omega\left(  \Theta\right)  }\delta\left[  \Theta-\hat{\Theta
}\right]  \equiv\frac{1}{\Omega\left(  I\right)  }\delta\left[  I-\hat
{I}\right]  . \label{ri}%
\end{equation}
The above identity (\ref{ri}) expresses an internal symmetry which will be
referred as a \textit{reparametrization invariance of the microcanonical
ensemble}. The reparametrization invariance of the microcanonical distribution
function $\hat{\omega}_{m}$ leads to the reparametrization invariance of the
expectation value of any microscopic quantity $\hat{A}$, $A=Sp\left\{  \hat
{A}\hat{\omega}_{m}\right\}  \Rightarrow A\left(  I\right)  \equiv A\left(
\Theta\right)  $, that is, the expectation values of the microscopic
observables behaves as scalar functions under any reparametrization change
$\Theta$.

Mathematically speaking, it is said that the Euclidean subsets $\mathcal{R}%
_{I}$ and $\mathcal{R}_{\Theta}$ are identical \textit{representations} of
certain abstract space $\Im$. The totality of the reparametrization changes
among the admissible representations of the space $\Im$ constitutes a group,
the \textit{Group of Diffeomorphisms} of the space $\Im$. The
reparametrization invariance of the microcanonical ensemble means that such
description provides the same macroscopic Physics for all admissible
representations of the abstract space $\Im$.

The microcanonical partition function $\Omega$ defines an invariant measure
$d\mu=\Omega\left(  \Theta\right)  d\Theta\equiv\Omega\left(  I\right)  dI$
which provides the number of microstates $W_{\sigma}=\int_{\pi_{\sigma}}d\mu
$\ \ belonging to a small subset $\pi_{\sigma}$ of certain \textit{coarsed
grained partition}
\begin{equation}
\mathcal{P}=\left\{  \pi_{\sigma}\subset\Im\left\vert
{\textstyle\bigcup\limits_{\sigma}}
\pi_{\sigma}=\Im\right.  \right\}
\end{equation}
\ of the abstract space $\Im$, and allows us to introduce the Boltzmann
entropy $S_{B}\left[  \pi_{\sigma}\right]  =\ln W_{\sigma}$. Hereafter, I will
consider that the number of macrostates $W_{\sigma}$ grows exponentially with
the increasing of the number $n$ of the degrees of freedom of the interest
system, so that the thermodynamic limit:
\begin{equation}
s=\lim_{n\rightarrow\infty}\frac{S_{B}\left[  \pi_{\sigma}\right]  }{n},
\end{equation}
exists and it is independent of the nature of the coarsed grained partition
$\pi_{\sigma}$ of the space $\Im$ whenever every subset $\pi_{\sigma}$ be
small. This hypothesis allows us to take the reduced entropy $s$ as a
\textit{continuous scalar function} defined on the space $\Im$. When the
interest system is large enough, the number of microstates $W$ can be
estimated in the practice by $W_{1}\simeq\Omega\left(  I\right)  \delta I_{0}%
$, where $\delta I_{0}$ is a small constant volume. The estimate $W_{1}$
differs from the one obtained in other representation $W_{2}\simeq
\Omega\left(  \Theta\right)  \delta\Theta_{0}$. However these estimations
should lead to the same reduced entropy in the thermodynamic limit
$n\rightarrow\infty$:%
\begin{align}
s  &  =\lim_{n\rightarrow\infty}\frac{\ln W_{2}}{n}=\lim_{n\rightarrow\infty
}\frac{\ln W_{1}}{n}\nonumber\\
&  \Leftrightarrow\lim_{n\rightarrow\infty}\frac{1}{n}\ln\left[  \left\vert
\frac{\partial\Theta\left(  I\right)  }{\partial I}\right\vert \frac{\delta
I_{0}}{\delta\Theta_{0}}\right]  =0,
\end{align}
which imposes the restriction that the Jacobian of the reparametrization can
not grow exponentially with $n$. I shall assume for practical purposes the
validity of the approximation $S\equiv S_{1}=\ln W_{1}\simeq S_{2}=\ln W_{2}$
when the interest system is large enough.

\section{Open system\label{open}}

Let us now consider an experimental setup where the interest system is put in
contact with certain external apparatus $\wp_{\eta}$ which controls its
macroscopic behavior in a way that the expectation values of the physical
quantities $\hat{\Theta}\equiv\Theta\left(  \hat{I}\right)  $ are kept fixed:%
\begin{equation}
\left\langle \hat{\Theta}^{k}\right\rangle =Sp\left\{  \Theta^{k}\left(
\hat{I}\right)  \hat{\omega}\right\}  .
\end{equation}
The equilibrium distribution function $\hat{\omega}$ of the interest system
under these conditions can be derived from the standard procedure of
maximization of the entropy:
\begin{equation}
S\left[  \hat{\omega}\right]  =-Sp\left\{  \hat{\omega}\ln\hat{\omega
}\right\}  ,
\end{equation}
which leads directly to the following \textit{generalized Boltzmann-Gibbs
distribution}:%
\begin{equation}
\hat{\omega}_{GC}\left(  \eta\right)  =\frac{1}{Z\left(  \eta\right)  }%
\exp\left\{  -\eta_{k}\Theta^{k}\left(  \hat{I}\right)  \right\}  .
\label{gce}%
\end{equation}
The Lagrange multipliers $\eta=\left\{  \eta_{k}\right\}  $ are the
\textit{canonical parameters} of the external apparatus $\wp_{\eta}$, which
will be referred hereafter as a \textit{generalized thermostat} or
\textit{generalized reservoir}. I shall show now that the above distribution
function provides a suitable generalization of the well-known Boltzmann-Gibbs
distributions:%
\begin{equation}
\hat{\omega}_{BG}\left(  \beta\right)  =\frac{1}{Z\left(  \beta\right)  }%
\exp\left\{  -\beta_{i}\hat{I}^{i}\right\}  , \label{bgd}%
\end{equation}
which is just a special case of the generalized distribution (\ref{gce}) with
$\Theta\left(  \hat{I}\right)  \equiv\hat{I}$. The convention $\lambda
_{k}\cdot\varphi^{k}\equiv\lambda\cdot\varphi$ will be hereafter adopted in
order to simplify our notation.

Since the operator:
\begin{equation}
\exp\left[  -\eta\cdot\Theta\left(  \hat{I}\right)  \right]  \equiv\int
\exp\left[  -\eta\cdot\Theta\left(  I\right)  \right]  \delta\left[  I-\hat
{I}\right]  dI,
\end{equation}
the canonical partition function $Z\left(  \eta\right)  $ can expressed as
follows:%
\begin{equation}
Z\left(  \eta\right)  =\int\exp\left[  -\eta\cdot\Theta\left(  I\right)
\right]  \Omega\left(  I\right)  dI, \label{p1}%
\end{equation}
where $\Omega\left(  I\right)  =Sp\left\{  \delta\left[  I-\hat{I}\right]
\right\}  $ is the microcanonical partition function in the representation
$\mathcal{R}_{I}$ of the space $\Im$.

The integral (\ref{p1}) when $\Theta\left(  I\right)  \equiv I$ is just the
\textit{Laplace transformation}:%
\begin{equation}
Z\left(  \beta\right)  =\int\exp\left[  -\beta\cdot I\right]  \Omega\left(
I\right)  dI, \label{p2}%
\end{equation}
which supports in the thermodynamic limit $n\rightarrow\infty$ the validity of
the well-known \textit{Legendre transformation}:%
\begin{equation}
P\left(  \beta\right)  =\inf_{I^{\ast}}\left\{  \beta\cdot I-S\left(
I\right)  \right\}  , \label{LT1}%
\end{equation}
between the Planck thermodynamic potential $P\left(  \beta\right)  =-\ln
Z\left(  \beta\right)  $ of the Boltzmann-Gibbs distribution (\ref{bgd}) and
the coarsed grained entropy $S\left(  I\right)  =\ln\left[  \Omega\left(
I\right)  \delta I_{0}\right]  $ of the microcanonical ensemble (\ref{mic}).
The identity (\ref{LT1}) represents that a macrostate of the Boltzmann-Gibbs
ensemble (\ref{bgd}) with canonical parameter $\beta$ is \textit{equivalent}
in the thermodynamic limit $n\rightarrow\infty$ to a macrostate of the
microcanonical ensemble (\ref{mic}) whose set $I^{\ast}$ represents the point
of global minimum of the functional:
\begin{equation}
P\left(  I;\beta\right)  =\beta\cdot I-S\left(  I\right)  ,
\end{equation}
within the representation $\mathcal{R}_{I}$ of the abstract space $\Im$, which
satisfies the conditions:
\begin{equation}
\beta_{i}=\frac{\partial S\left(  I^{\ast}\right)  }{\partial I^{i}}%
,~\kappa_{ij}=\frac{\partial^{2}S\left(  I^{\ast}\right)  }{\partial
I^{i}\partial I^{j}}, \label{eq1}%
\end{equation}
where the \textit{entropy Hessian} $\kappa_{ij}$ must be a \textit{negative
defined matrix}, that is, the entropy must be locally concave in the point
$I^{\ast}\in\mathcal{R}_{I}$. Such exigency implies that all those regions of
the Euclidean subset $\mathcal{R}_{I}$ where the entropy $S$\ is not locally
concave will never be equivalent to a macrostate of the Boltzmann-Gibbs
ensemble (\ref{bgd}). This situation is usually referred as an
\textit{ensemble inequivalence}.

Let $\mathcal{B}_{I}$ be the Euclidean n-dimensional subset composed by all
admissible values of the canonical parameters $\beta$ of the external
apparatus leading to a thermodynamic equilibrium of the interest system
characterized by the Boltzmann-Gibbs distribution (\ref{bgd}). According to
the condition\ (\ref{eq1}), such thermodynamic equilibrium is characterized by
the identification $\tilde{\beta}=\hat{\beta}$ of the canonical parameters
$\tilde{\beta}$ of the external apparatus with the components $\hat{\beta}$ of
the gradient of the microcanonical entropy, the map $\psi_{I}$:
\begin{equation}
\psi_{I}:\mathcal{R}_{I}\rightarrow\mathcal{B}_{I}\equiv\left\{  \hat{\beta
}\in\mathcal{B}_{I}\left\vert \hat{\beta}_{i}=\frac{\partial S\left(
I\right)  }{\partial I^{i}}\right.  \right\}  ,
\end{equation}
in the representation $\mathcal{R}_{I}$ of the space $\Im$. When the entropy
is locally concave in certain subset $\pi_{\alpha}\subset\mathcal{R}_{I}$ ,
the map $\psi_{I}$ establishes a \textit{bijective} \textit{correspondence}
between the subset $\pi_{\alpha}$ and certain subset $\Lambda_{\alpha}%
\subset\mathcal{B}_{I}$. During the ensemble inequivalence the bijective
character of the map $\psi$ is lost. The set $\tilde{\beta}\in\mathcal{B}_{I}$
of canonical parameters of the external apparatus leads to a macrostate with
ensemble inequivalence when there exist at least two distinct points $I_{1}$
and $I_{2}\in\mathcal{R}_{I}$, where $\tilde{\beta}=\psi_{I}\left(
I_{1}\right)  =\psi_{I}\left(  I_{2}\right)  $. Thus, the ensemble
inequivalence can be seen as a thermodynamic equilibrium where exist a
\textit{competition} among all those microcanonical states $I_{k}%
\in\mathcal{R}_{I}$ where $\psi_{I}\left(  I_{k}\right)  =\tilde{\beta}$.

Let us return to the case of the generalized Boltzmann-Gibbs ensemble
(\ref{gce}). Taking into account the reparametrization invariance of the
measure $d\mu=\Omega\left(  I\right)  dI=\Omega\left(  \Theta\right)  d\Theta
$, the integral (\ref{p1}) can be rephrased as follows%
\begin{equation}
Z\left(  \eta\right)  =\int\exp\left[  -\eta\cdot\Theta\right]  \Omega\left(
\Theta\right)  d\Theta,
\end{equation}
which allows a natural extension of the conventional Statistical Mechanics and
Thermodynamics with a simple reparametrization change $I\rightarrow\Theta$ and
$\beta\rightarrow\eta$: The Legendre transformation between the thermodynamic
potentials $P\left(  \eta\right)  =-\ln Z\left(  \eta\right)  $ and $S\left(
\Theta\right)  =\ln\left[  \Omega\left(  \Theta\right)  \delta\Theta
_{0}\right]  $:
\begin{equation}
P\left(  \eta\right)  =\inf_{\Theta^{\ast}}\left\{  \eta\cdot\Theta-S\left(
\Theta\right)  \right\}  , \label{LT2}%
\end{equation}
and the conditions of the ensemble equivalence:%
\begin{equation}
\eta_{k}=\frac{\partial S\left(  \Theta^{\ast}\right)  }{\partial\Theta^{k}%
},~\kappa_{km}=\frac{\partial^{2}S\left(  \Theta^{\ast}\right)  }%
{\partial\Theta^{k}\partial\Theta^{m}}, \label{eq2}%
\end{equation}
where the entropy Hessian $\kappa_{km}$ in the representation $\mathcal{R}%
_{\Theta}$ must be a negative definite matrix, that is, the entropy $S\left(
\Theta\right)  $ must be locally concave in the point $\Theta^{\ast}\in$
$\mathcal{R}_{\Theta}$. The above results deserve some remarks.

\begin{remark}
While a reparametrization change does not alter the microcanonical description
of an isolated system, for an open system such transformation implies a
substitution of the external apparatus \ $\wp_{\beta}\rightarrow\wp_{\eta}$
which controls its thermodynamic equilibrium, and consequently, generalized
Boltzmann-Gibbs ensembles with different representation $\mathcal{R}%
_{\Theta_{1}}$ and $\mathcal{R}_{\Theta_{2}}$ provide in general different
thermodynamical descriptions.
\end{remark}

\begin{remark}
The validity of the Legendre transformation (\ref{LT2}) implies the local
equivalence of the macrostates of different generalized Boltzmann-Gibbs
ensembles in the thermodynamic limit $n\rightarrow\infty$ whenever they are
equivalent to the same macrostate of the microcanonical ensemble. Such
equivalence implies a relation among the canonical parameters of the
generalized thermostat and an ordinary thermostat: Since $S\left(  I\right)
\equiv S\left(  \Theta\right)  $ when $n\rightarrow\infty$, the conditions
(\ref{eq1}) and (\ref{eq2}) lead the following transformation rule during a
reparametrization change $\Theta\rightarrow I$:
\begin{equation}
\left.
\begin{array}
[c]{c}%
\beta_{i}=\partial S/\partial I^{i}\\
\eta_{k}=\partial S/\partial\Theta^{k}%
\end{array}
\right\}  \Rightarrow\beta_{i}=\frac{\partial\Theta^{k}}{\partial I^{i}}%
\eta_{k}, \label{tr1}%
\end{equation}
while the entropy Hessian obeys:%
\begin{equation}
\kappa_{ij}=\frac{\partial\Theta^{k}}{\partial I^{i}}\frac{\partial\Theta^{m}%
}{\partial I^{j}}\kappa_{km}+\frac{\partial^{2}\Theta^{k}}{\partial
I^{i}\partial I^{j}}\eta_{k}. \label{tr2}%
\end{equation}

\end{remark}

Since $\mathcal{R}_{I}$ and $\mathcal{R}_{\Theta}$ are just two admissible
representations of the abstract space $\Im$, the above transformation rules
are also applicable to any two admissible representations of $\Im$. Equation
(\ref{tr1}) is just the transformation rule of the components of a covariant
vector within a differential geometry. However, the entropy Hessian does not
represent a second rank covariant tensor, since the transformation rule should
be given by:
\begin{equation}
\tau_{ij}=\frac{\partial\Theta^{k}}{\partial I^{i}}\frac{\partial\Theta^{m}%
}{\partial I^{j}}\tau_{km}%
\end{equation}
instead of the rule (\ref{tr2}). This result leads to an important third remark.

\begin{remark}
Since the entropy Hessian is not a second rank covariant tensor, the concavity
properties of the entropy may change during the reparametrization changes.
Consequently, the regions of ensemble inequivalence between the
Boltzmann-Gibbs ensemble and the microcanonical ensemble in the representation
$\mathcal{R}_{I}$ does not coincide to the regions of ensemble inequivalence
in the representation $\mathcal{R}_{\Theta}$.
\end{remark}

Let us consider a trivial example taken from our previous work \cite{vel.geo}.
Let $s$ be a positive real map defined on a seminfinite Euclidean line
$\mathcal{L}$, $s:\mathcal{L}\rightarrow R^{+}$, which is given by the concave
function $s\left(  x\right)  =\sqrt{x}$ in the representation $\mathcal{R}%
_{x}$ of\ $\mathcal{L}$ (where $x>0$). Let $\varphi$ be a reparametrization
change $\varphi:\mathcal{R}_{x}\rightarrow\mathcal{R}_{y}$ given by
$y=\varphi\left(  x\right)  =x^{\frac{1}{4}}$. The map $s$ in the new
representation $\mathcal{R}_{y}$ of the seminfinite Euclidean line
$\mathcal{L}$ is now given by the function $s\left(  y\right)  =y^{2}$ (with
$y>0$), which is clearly a convex function.

The third remark implies a revision of our conceptions about the
classification of the phase transitions based on the concavity of the entropy
and the generalization of some Monte Carlo methods with distribution functions
inspired on the Statistical Mechanics.

The first-order phase transitions are closely related with the phenomenon of
ensemble inequivalence. Since the ensemble inequivalence depends on the nature
of the external apparatus controlling the thermodynamic equilibrium of the
interest system, \textit{the existence of this anomaly reflects the inability
of the external apparatus to control all microcanonically admissible
equilibrium states of the interest system, and consequently, this kind of
phenomenon can not be relevant within the microcanonical description}. The
interested reader can see a long discussion about this subject in the
ref.\cite{vel.geo}.

The phenomenon of ensemble inequivalence also implies \textit{an important the
lost of information }about the thermodynamical properties of a given system
during the phase coexistence in regard to the thermodynamical information
which can be derived from the microcanonical description. Such lost of
information also leads to the failure of ordinary Monte Carlo methods based on
the Boltzmann-Gibbs ensemble in the neighborhood of the first-order phase
transitions \cite{wang2}.

The ensemble inequivalence can be successfully avoided by using an appropriate
representation $\mathcal{R}_{\Theta}$ within the generalized Boltzmann-Gibbs
distribution (\ref{gce}). Therefore, the simple consideration of the
reparametrization changes allows a suitable extension of some Monte Carlo
methods. This question was discussed in the ref.\cite{vel-mmc} for the
particular case of the Gibbs canonical ensemble. I shall consider below a
formal extension of the methodology developed in that reference for enhancing
the well-known Metropolis importance sample algorithm \cite{met}\ by focussing
only the most important features. The reader could see the ref.\cite{vel-mmc}
for more details.

The probability of acceptance of a Metropolis move based on the generalized
Boltzmann-Gibbs distribution (\ref{gce}) can be given by:
\begin{equation}
p\left(  \left.  I\right\vert I+\Delta I\right)  =\min\left\{  1,\exp\left[
-\eta\cdot\Delta\Theta\right]  \right\}  ,
\end{equation}
where $\Delta\Theta^{k}=\Theta^{k}\left(  I+\Delta I\right)  -\Theta\left(
I\right)  $. When $n$ is large enough $\Delta I<<I\Rightarrow\Delta\Theta
^{k}\simeq\left\{  \partial\Theta^{k}\left(  I\right)  /\partial
I^{i}\right\}  \Delta I^{i}$, and consequently:
\begin{equation}
p\left(  \left.  I\right\vert I+\Delta I\right)  \cong\min\left\{
1,\exp\left[  -\tilde{\beta}\cdot\Delta I\right]  \right\}  ,
\end{equation}
where the effective canonical parameter $\tilde{\beta}_{i}=\beta_{i}\left(
I;\eta\right)  $ is given by:%
\begin{equation}
\tilde{\beta}_{i}=\beta_{i}\left(  I;\eta\right)  =\frac{\partial\Theta
^{k}\left(  I\right)  }{\partial I^{i}}\eta_{k}. \label{fluc}%
\end{equation}

Reader can notice that the above definition is very similar to the
transformation rule (\ref{tr1}). The difference resides now in the fact that
the set $I$ does not take the equilibrium value $I^{\ast}$, but it changes
during the Monte Carlo dynamics. Thus, the present generalized Metropolis
algorithm looks-like an ordinary Metropolis algorithm where \textit{the
canonical parameters }$\tilde{\beta}=\beta\left(  I;\eta\right)  $\textit{ of
the external reservoir exhibits correlated fluctuations with the fluctuations
of the physical quantities }$I$\textit{ characterizing the interest system
around the corresponding equilibrium values }$\beta$ and $I^{\ast}$\textit{
connected by the ensemble equivalence within the generalized Boltzmann-Gibbs
description }(\ref{gce})\textit{ and the transformation rule }(\ref{tr1}). As
already shown in ref.\cite{vel-mmc}, this feature of the present Metropolis
algorithm allows the Monte Carlo dynamics to explore the anomalous regions of
the subset \ $\mathcal{R}_{I}$\ where the ensemble inequivalence within the
Boltzmann-Gibbs description (\ref{bgd}) takes place. The reader can see an
example of application of this method in the section \ref{apply}. The above
observations lead directly to a fourth remark which clarifies us the nature of
the generalized thermostat $\wp_{\eta}$.

\begin{remark}
When the interest system is large enough, the generalized thermostat or
reservoir $\wp_{\eta}$ leading to the generalized Boltzmann-Gibbs distribution
(\ref{gce}) is almost equivalent to an ordinary thermostat $\wp_{\tilde{\beta
}}$\ whose canonical parameters $\tilde{\beta}$ fluctuates with the
fluctuations of the physical quantities $I$ of the interest system according
to the rule (\ref{fluc}).
\end{remark}

The fourth remark is particularly interesting. Conventional Thermodynamics
usually deals with equilibrium situations where the fundamental physical
quantities $I$ are kept fixed when the interest system is isolated (a
microcanonical description) or the physical quantities $I$ fluctuate and
corresponding canonical parameters $\beta$ are kept fixed by the thermostat
("within canonical description"). The generalized Boltzmann-Gibbs ensemble
(\ref{gce}) is just a natural framework for consider all those equilibrium
situations where both the fundamental physical quantities $I$ of the interest
system and the corresponding effective canonical parameters $\tilde{\beta}$ of
the external thermostat fluctuate as a consequence of their mutual
interaction. Thus, the ordinary condition for the thermodynamic equilibrium
between the canonical parameters of the interest system $\hat{\beta}$ and the
external apparatus $\tilde{\beta}$ only takes place for their corresponding
average values:%
\begin{equation}
\left\langle \tilde{\beta}\right\rangle =\left\langle \hat{\beta}\right\rangle
.
\end{equation}
The last observation leads directly to the main objective of the present
study: the implementation of a generalized fluctuation theory based on the
reparametrization invariance of the microcanonical ensemble.

\section{Generalized Fluctuation Theory}

\subsection{Fundamental results}

Let $\delta I=I-I^{\ast}$ and $\delta\beta=\tilde{\beta}-\beta$ be dispersions
of the fundamental physical quantities $I$ of the interest system and the
effective canonical parameters $\tilde{\beta}$ of the generalized thermostat
respectively. Our aim is to obtain the correlations $\left\langle \delta
I^{i}\delta I^{j}\right\rangle $, $\left\langle \delta\beta_{i}\delta
I^{j}\right\rangle $ and $\left\langle \delta\beta_{i}\delta\beta
_{j}\right\rangle $ within the generalized Boltzmann-Gibbs description
(\ref{gce}) and identify their mutual relationships. It is very important to
remark that such correlations depends crucially on the interaction between the
interest system and the generalized thermostat, and consequently, the values
of these quantities are modified with the change of the external control apparatus.

Let us began from the correlation $\left\langle \delta I^{i}\delta
I^{j}\right\rangle $:
\begin{equation}
\left\langle \delta I^{i}\delta I^{j}\right\rangle =\frac{1}{Z\left(
\eta\right)  }\int\delta I^{i}\delta I^{j}\exp\left[  -\eta\cdot\Theta\left(
I\right)  \right]  \Omega\left(  I\right)  dI,
\end{equation}
where the Gaussian approximations leads directly to the result:%
\begin{equation}
\left\langle \delta I^{i}\delta I^{j}\right\rangle =\left(  \Pi_{ij}^{\left(
I\right)  }\text{ }\right)  ^{-1}. \label{pi0}%
\end{equation}
where the matrix $\Pi_{ij}^{\left(  I\right)  }$ is given by:
\begin{equation}
\Pi_{ij}^{\left(  I\right)  }=\eta_{k}\frac{\partial^{2}\Theta\left(  I^{\ast
}\right)  }{\partial I^{i}\partial I^{j}}-\frac{\partial^{2}S\left(  I^{\ast
}\right)  }{\partial I^{i}\partial I^{j}}. \label{pi1}%
\end{equation}
The above result generalizes the classical relation among the correlations
$\left\langle \delta I^{i}\delta I^{j}\right\rangle $ and the entropy Hessian
$\kappa_{ij}$ (\ref{eq1}), $\left\langle \delta I^{i}\delta I^{j}\right\rangle
=-$ $\kappa_{ij}^{-1}$ derived from the Boltzmann-Gibbs distribution
(\ref{bgd}). Taking into consideration the transformation rule (\ref{tr2}),
the matrix $\Pi_{ij}^{\left(  I\right)  }$ can be rewritten as follows:%
\begin{equation}
\Pi_{ij}^{\left(  I\right)  }\equiv-\frac{\partial\Theta^{k}}{\partial I^{i}%
}\frac{\partial\Theta^{m}}{\partial I^{j}}\kappa_{km}, \label{n1}%
\end{equation}
where $\kappa_{km}$ is the entropy Hessian (\ref{eq2}) in the representation
$\mathcal{R}_{\Theta}$ of the abstract space $\Im$.

The expectation values $\Theta^{\ast}=\left\langle \Theta^{k}\right\rangle $
and $\left\langle \delta\Theta^{k}\delta\Theta^{m}\right\rangle $ with
$\delta\Theta=\Theta-\Theta^{\ast}$ within the generalized Boltzmann-Gibbs
description (\ref{gce}) can be obtained from the Planck thermodynamic
potential $P\left(  \eta\right)  $ in the representation $\mathcal{R}_{\Theta
}$ in complete analogy with the standard Statistical Mechanics:%

\begin{equation}
\left\langle \Theta^{k}\right\rangle =\frac{\partial P\left(  \eta\right)
}{\partial\eta_{k}},~\left\langle \delta\Theta^{k}\delta\Theta^{m}%
\right\rangle =-\frac{\partial^{2}P\left(  \eta\right)  }{\partial\eta
_{k}\partial\eta_{m}}.
\end{equation}
The Gaussian estimation of $\left\langle \delta\Theta^{k}\delta\Theta
^{m}\right\rangle $ is given now by the inverse of the entropy Hessian
(\ref{eq2}):
\begin{equation}
\left\langle \delta\Theta^{k}\delta\Theta^{m}\right\rangle =-\kappa_{km}%
^{-1}=\left(  \Pi_{km}^{\left(  \Theta\right)  }\right)  ^{-1}.
\end{equation}
Taking into account the result (\ref{n1}):%

\begin{equation}
\Pi_{ij}^{\left(  I\right)  }=\frac{\partial\Theta^{k}}{\partial I^{i}}%
\frac{\partial\Theta^{m}}{\partial I^{j}}\Pi_{km}^{\left(  \Theta\right)  },
\end{equation}
and therefore, the correlations $\left\langle \delta I^{i}\delta
I^{j}\right\rangle $ and $\left\langle \delta\Theta^{k}\delta\Theta
^{m}\right\rangle $ are related by a transformation rule of second rank
contravariant tensors:
\begin{equation}
\left\langle \delta I^{i}\delta I^{j}\right\rangle =\frac{\partial I^{i}%
}{\partial\Theta^{k}}\frac{\partial I^{j}}{\partial\Theta^{m}}\left\langle
\delta\Theta^{k}\delta\Theta^{m}\right\rangle ,
\end{equation}
where the correlations $\left\langle \delta I^{i}\delta I^{j}\right\rangle $
can be expresses in terms of the Planck thermodynamic potential $P\left(
\eta\right)  $ as follows:
\begin{equation}
\left\langle \delta I^{i}\delta I^{j}\right\rangle =-\frac{\partial I^{i}%
}{\partial\Theta^{k}}\frac{\partial I^{j}}{\partial\Theta^{m}}\frac
{\partial^{2}P\left(  \eta\right)  }{\partial\eta_{k}\partial\eta_{m}}.
\end{equation}

The correlations $\left\langle \delta\beta_{i}\delta I^{j}\right\rangle $ and
$\left\langle \delta\beta_{i}\delta\beta_{j}\right\rangle $ can be derived
easily from the correlations $\left\langle \delta I^{i}\delta I^{j}%
\right\rangle $. Since $\delta I<<I^{\ast}$ when the interest system is large
enough, the dispersion $\delta\beta$ of the effective canonical parameters
$\tilde{\beta}=\beta\left(  I;\eta\right)  $ defined in the equation
(\ref{fluc}) can be estimated in terms of the dispersions $\delta I$ as
follows:
\begin{equation}
\delta\beta_{i}\cong\eta_{k}\frac{\partial^{2}\Theta^{k}\left(  I^{\ast
}\right)  }{\partial I^{i}\partial I^{j}}\delta I^{j}, \label{appb}%
\end{equation}
and therefore:%
\begin{equation}
\left\langle \delta\beta_{i}\delta I^{j}\right\rangle =\eta_{k}\frac
{\partial^{2}\Theta^{k}\left(  I^{\ast}\right)  }{\partial I^{i}\partial
I^{n}}\left\langle \delta I^{n}\delta I^{j}\right\rangle \equiv F_{in}%
\left\langle \delta I^{n}\delta I^{j}\right\rangle . \label{id2}%
\end{equation}
Since the definition (\ref{pi1}) of the matrix $\Pi_{ij}^{\left(  I\right)  }$
allows us to express the matrix $F_{ij}$ derived from the second derivatives
of the map $\Theta$ as follows:
\begin{equation}
F_{ij}\equiv\eta_{k}\frac{\partial^{2}\Theta^{k}\left(  I^{\ast}\right)
}{\partial I^{i}\partial I^{j}}=\Pi_{ij}^{\left(  I\right)  }+\frac
{\partial^{2}S\left(  I^{\ast}\right)  }{\partial I^{i}\partial I^{j}},
\end{equation}
the correlations $\left\langle \delta\beta_{i}\delta I^{j}\right\rangle $ can
be given by:%
\begin{equation}
\left\langle \delta\beta_{i}\delta I^{j}\right\rangle =\delta_{i}^{j}%
+\frac{\partial^{2}S\left(  I^{\ast}\right)  }{\partial I^{i}\partial I^{n}%
}\left\langle \delta I^{n}\delta I^{j}\right\rangle , \label{gen1}%
\end{equation}
where the delta Kronecker $\delta_{i}^{j}$ appears as consequence of the
presence of the terms $\Pi_{in}^{\left(  I\right)  }\left\langle \delta
I^{n}\delta I^{j}\right\rangle \equiv\delta_{i}^{j}$ which is
straightforwardly followed from the equation (\ref{pi0}).

Finally, the approximation (\ref{appb}) and the identity (\ref{id2}) allows us
to express the correlations $\left\langle \delta\beta_{i}\delta\beta
_{j}\right\rangle $ as follows:
\begin{equation}
\left\langle \delta\beta_{i}\delta\beta_{j}\right\rangle =F_{in}%
F_{jm}\left\langle \delta I^{n}\delta I^{m}\right\rangle =F_{jm}\left\langle
\delta\beta_{i}\delta I^{m}\right\rangle ,
\end{equation}
and considering the relation:\qquad%
\begin{equation}
F_{jm}\equiv\left\langle \delta\beta_{j}\delta I^{n}\right\rangle \Pi
_{nm}^{\left(  I\right)  },
\end{equation}
I arrive to the following result:%
\begin{equation}
\left\langle \delta\beta_{i}\delta\beta_{j}\right\rangle =\left\langle
\delta\beta_{j}\delta I^{n}\right\rangle \Pi_{nm}^{\left(  I\right)
}\left\langle \delta\beta_{i}\delta I^{m}\right\rangle . \label{gen2}%
\end{equation}

The relations (\ref{gen1}) and (\ref{gen2}) are the fundamental results of the
present subsection. Introducing the nomenclature for the correlations tensors
$G^{ij}=\left\langle \delta I^{i}\delta I^{j}\right\rangle $, $M_{i}%
^{j}=\left\langle \delta\beta_{i}\delta I^{j}\right\rangle $ and
$T_{ij}=\left\langle \delta\beta_{i}\delta\beta_{j}\right\rangle $, the
fundamentals relations can be rewritten as follows:%
\begin{align}
M_{i}^{j}  &  =\delta_{i}^{j}+\kappa_{im}G^{mj},\label{a1}\\
T_{ij}  &  =M_{j}^{n}G_{nm}M_{i}^{m}, \label{a2}%
\end{align}
where $G_{nm}=\left(  G^{-1}\right)  _{nm}\equiv\Pi_{nm}^{\left(  I\right)  }$.

I can substitute (\ref{a1}) in (\ref{a2}) in order to express the correlation
matrix $M_{i}^{j}$ in terms of the entropy Hessian $\kappa_{ij}$ and the
correlation matrix $G^{ij}$:%
\begin{align}
T_{ij}  &  =M_{j}^{n}G_{nm}\left(  \delta_{i}^{m}+\kappa_{il}G^{lm}\right)
\nonumber\\
&  =M_{j}^{n}\left(  G_{ni}+\kappa_{in}\right)  ,\nonumber\\
&  =\left(  \delta_{j}^{n}+\kappa_{jm}G^{mn}\right)  \left(  G_{ni}%
+\kappa_{in}\right)  ,\nonumber\\
&  =G_{ij}+2\kappa_{ij}+\kappa_{in}\kappa_{jm}G^{nm}, \label{r3}%
\end{align}
which can be rewritten as follows:%
\begin{align}
G^{\alpha i}T_{ij}  &  =\delta_{j}^{\alpha}+2G^{\alpha i}\kappa_{ij}+G^{\alpha
i}\kappa_{in}G^{nm}\kappa_{jm}\nonumber\\
&  =\left(  \delta_{\beta}^{\alpha}+G^{\alpha i}\kappa_{i\beta}\right)
\left(  \delta_{i}^{\beta}+G^{\beta n}\kappa_{ni}\right)  ,\nonumber\\
&  =M_{n}^{\alpha}M_{j}^{n}.
\end{align}
Therefore, the identity (\ref{a2}) can be expressed as follows%
\begin{equation}
\left\langle \delta I^{i}\delta I^{n}\right\rangle \left\langle \delta
\beta_{n}\delta\beta_{j}\right\rangle =\left\langle \delta I^{i}\delta
\beta_{n}\right\rangle \left\langle \delta I^{n}\delta\beta_{j}\right\rangle .
\end{equation}

The reader may notice that no one of the above identities makes any explicit
reference to the representation $\mathcal{R}_{\Theta}$ used in the generalized
Boltzmann-Gibbs ensemble (\ref{gce}), and consequently, such thermodynamical
identities possesses a general validity for the case of an external apparatus
controlling the average values of the fundamental quantities $I$ of the
interest system throughout the identification of the average values of the
effective canonical parameters $\tilde{\beta}$ of the external apparatus with
the corresponding average values of the parameters $\hat{\beta}=\partial
S/\partial I$ derived from the gradient of the microcanonical entropy $S$ of
the interest system.

Since the thermodynamical variables $I$ and $\beta$ are just an admissible
representation from the reparametrization invariance viewpoint, the
reparametrization change $I\rightarrow\Theta$ and $\beta\rightarrow\eta$ leads
also to the validity of the relations:%
\begin{align}
\left\langle \delta\eta_{k}\delta\Theta^{m}\right\rangle  &  =\delta_{k}%
^{m}+\frac{\partial^{2}S\left(  \Theta\right)  }{\partial\Theta^{k}%
\partial\Theta^{n}}\left\langle \delta\Theta^{n}\delta\Theta^{m}\right\rangle
,\\
\left\langle \delta\Theta^{r}\delta\Theta^{m}\right\rangle \left\langle
\delta\eta_{m}\delta\eta_{k}\right\rangle  &  =\left\langle \delta\eta
_{k}\delta\Theta^{n}\right\rangle \left\langle \delta\eta_{n}\delta\Theta
^{r}\right\rangle ,
\end{align}
which means that the fundamental results also exhibit reparametrization invariance.

\subsection{Some anomalous behaviors revised}

As elsewhere shown \cite{rupper}, the entropy Hessian provides important
information about the fluctuations of thermodynamic quantities during the
thermodynamic equilibrium of an open system. Such connection is usually
derived from the Boltzmann-Gibbs distribution (\ref{bgd}) which leads directly
within the Gaussian approximation to the following result:%
\begin{equation}
\kappa_{ij}=\frac{\partial^{2}S}{\partial I^{i}\partial I^{j}}\rightarrow
-\kappa_{im}\left\langle \delta I^{m}\delta I^{j}\right\rangle =\delta_{i}%
^{j}. \label{HG1}%
\end{equation}
The correlation matrix $G^{ij}=\left\langle \delta I^{i}\delta I^{j}%
\right\rangle $ is always nonnegative, which is easily verified from the
following reasonings:%
\begin{equation}
\left\langle \left(  \xi_{i}\delta I^{i}\right)  ^{2}\right\rangle =\xi_{i}%
\xi_{i}\left\langle \delta I^{i}\delta I^{j}\right\rangle \equiv\xi_{i}%
G^{ij}\xi_{j}\geq0,
\end{equation}
where the equality only takes place when $\xi\equiv0$. Consequently, the
applicability of the relation (\ref{HG1}) is limited to those regions of the
subset $\mathcal{R}_{I}$ where the entropy Hessian be a negative definite
matrix, that is, the regions in which is ensured the equivalence between the
Boltzmann-Gibbs ensemble (\ref{bgd}) and the microcanonical ensemble. Thus,
the extrapolation of the formula (\ref{HG1}) towards regions of ensemble
inequivalence has no sense within the conventional Thermodynamics. Let us
consider a very simple example.

Let be a system whose thermodynamical state is determined only by the total
energy $E$. The corresponding canonical parameter is the inverse temperature
$\beta$ of the Gibbs thermostat. The relation (\ref{HG1}) is given now by:%
\begin{equation}
-\frac{\partial^{2}S\left(  E\right)  }{\partial E^{2}}\left\langle \delta
E^{2}\right\rangle =1,
\end{equation}
which can be conveniently rewritten by using the microcanonical inverse
temperature $\beta\left(  E\right)  =1/T\left(  E\right)  =\partial S\left(
E\right)  /\partial E$ as follows:
\begin{align}
\left\langle \delta E^{2}\right\rangle  &  =-\left(  \frac{\partial
^{2}S\left(  E\right)  }{\partial E^{2}}\right)  ^{-1}=-\left(  \frac
{\partial\beta\left(  E\right)  }{\partial E}\right)  ^{-1}\nonumber\\
&  =T^{2}\frac{dE}{dT}=T^{2}C, \label{squarece}%
\end{align}
where $C$ is the heat capacity. Since\ the average square dispersion of the
total energy energy $\left\langle \delta E^{2}\right\rangle \geq0$, the heat
capacity should be nonnegative $C\geq0$. As already mentioned, such relation
is only applicable during the ensemble inequivalence.

There is nothing wrong in assuming that all those energetic regions with\ a
convex entropy $\partial^{2}S\left(  E\right)  /\partial E^{2}>0$ are just
thermodynamical states with a \textit{negative heat capacity} $C<0$. Such
anomalous behaviors are well-known since the famous Lyndel-Bell work
\cite{Lynden} and them have been also observed recently in experiments
involving nuclear or molecular clusters \cite{moretto,Dagostino,gro na}.
However, the relation between the heat capacity $C$ with the average energy
fluctuations (\ref{squarece}) has to be generalized in order to attribute some
reasonable physical meaning to these anomalous thermodynamical states in terms
of a suitable fluctuation theory. Such generalization is provided precisely by
the formula:%
\begin{equation}
\left\langle \delta\beta\delta E\right\rangle =1+\frac{\partial^{2}S\left(
E\right)  }{\partial E^{2}}\left\langle \delta E^{2}\right\rangle ,
\end{equation}
which is a particular expression of the fundamental result (\ref{gen1}) to the
unidimensional case (only one control parameter). While the inverse
temperature can be kept\ fixed $\delta\beta=0$ in order to control a
microcanonical state with a positive heat capacity $C>0$, such control
parameter can not be kept fixed when we deal with microcanonical states with
non positive heat capacities $C\leq0$. There $\left\langle \delta\beta\delta
E\right\rangle \geq1$, which means that the fluctuations $\beta$ and $I$
should be correlated in order to ensure the control of such anomalous
thermodynamic states, that is, in order to ensure the ensemble equivalence. As
already commented in the subsection above, this feature is the key for the
success of the Metropolis algorithm based on the generalized Boltzmann-Gibbs
ensemble (\ref{gce}).

The heat capacity $C$ is a very simple example of a \textit{response function
}which characterizes how sensible is the system energy $E$ (controlled
quantity) during a small change of the temperature $T$ of the thermostat
(control parameter). Another example of response function is the magnetic
susceptibility $\chi_{B}=\partial M/\partial B$, which characterizes how
sensible is the system magnetization $M$ (controlled quantity) during a small
change of the external magnetic field $B$ (control parameter). Thus, the
well-known thermodynamical identities:%
\begin{equation}
C=\frac{\partial E}{\partial T}=\beta^{2}\left\langle \delta E^{2}%
\right\rangle ,~\chi_{B}=\frac{\partial M}{\partial B}=\beta\left\langle
\delta M^{2}\right\rangle , \label{usual RF}%
\end{equation}
can be considered as \textit{response-fluctuation theorems}. As already
pointed out, such relations are non applicable during the ensemble equivalence
where $C$ or $\chi_{B}$ can admits negative values, and therefore, such
thermodynamic theorems should find a natural extension within the present framework.

\subsection{Microcanonical thermodynamic formalism\label{formalism}}

The geometric framework developed in the present work constitutes a natural
generalization of the thermodynamic formalism of the conventional
Thermodynamics. A key for the success is to perform a thermodynamical
description whose essence be so close to the macroscopic picture provided by
the microcanonical ensemble, which possesses a hierarchical supremacy in
regard to any generalized Boltzmann-Gibbs description (\ref{gce}).

The total differentiation of the microcanonical entropy:
\begin{equation}
dS=\beta_{i}dI^{i}~\text{with }\beta_{i}=\frac{\partial S}{\partial I^{i}},
\label{total_dif}%
\end{equation}
provides the expectation values of the effective canonical parameters
$\tilde{\beta}$ of an external generalized thermostat and the physical
quantities $I$ in the representation $\mathcal{R}_{I}$ of the abstract space
$\Im$. The symmetry of the entropy Hessian $\kappa_{ij}$ leads to what could
be considered as the \textit{Maxwell identities} within the microcanonical
description:
\begin{equation}
\kappa_{ij}=\frac{\partial^{2}S}{\partial I^{i}\partial I^{j}}=\kappa
_{ji}\Rightarrow\frac{\partial\beta_{i}}{\partial I^{j}}\equiv\frac
{\partial\beta_{j}}{\partial I^{i}}. \label{maxwell}%
\end{equation}

The reparametrization invariance imposes some restrictions to the admissible
parameters used in the generalized Boltzmann-Gibbs canonical description
within the present geometric framework. As already shown in the equation
(\ref{tr1}), the canonical parameters $\beta_{i}$ are just the components of a
covariant vector. Such status can not be attributed to the ordinary
temperature $T$ or the external magnetic field $B$. For example, the Gibbs
canonical ensemble written for a ferromagnetic system with Hamiltonian
$\hat{H}$ under an external magnetic field $B$ directed along the z-axis:
\begin{equation}
\hat{\omega}_{G}=\frac{1}{Z\left(  \beta;B_{z}\right)  }\exp\left[
-\beta\left(  \hat{H}-B\hat{M}_{z}\right)  \right]  , \label{ex1}%
\end{equation}
can be considered as a Boltzmann-Gibbs distribution with canonical parameters
$\beta$ and $\lambda=-\beta B$, whose microcanonical description is given by:%
\begin{equation}
\hat{\omega}_{M}=\frac{1}{\Omega\left(  U,M_{z}\right)  }\delta\left(
U-\hat{H}\right)  \delta\left(  M_{z}-\hat{M}_{z}\right)  . \label{ex2}%
\end{equation}
While the parametrization $\left(  \beta,\lambda\right)  $ is just a covariant
vector within a differential geometry, such character is not exhibit by the
the pair $\left(  T,B\right)  $, and therefore, this last is not a suitable
parametrization of the generalized canonical description from the
reparametrization invariance point of view. The above observation inspires a
reasonable and necessary \textit{redefinition of the concept of response
functions in order to preserve the geometric covariance of the thermodynamic
formalism developed in this work}.

The element of the \textit{generalized response matrix} $\chi^{ij}$:
\begin{equation}
\chi^{ij}=-\frac{\partial\left\langle I^{i}\right\rangle }{\partial\beta_{j}%
}\text{,} \label{response matrix}%
\end{equation}
characterizes the rate of change of the average of the \textit{i-th} physical
quantity $\left\langle I^{i}\right\rangle $ under a small variation of the
\textit{j-th} average effective control parameter $\beta_{j}=\left\langle
\tilde{\beta}_{j}\right\rangle $ of the generalized thermostat. Thus,
$\chi^{ij}$ is a measure of the sensibility of the interest system under the
control of the external apparatus. Strictly speaking, the response matrix
$\chi^{ij}$ (\ref{response matrix}) is not a second-rank contravariant tensor.
Such character is straightforwardly followed from the fact that
the\ microcanonical counterpart $\hat{\chi}_{ij}$\ of the \textit{inverse
response matrix }$\chi_{ij}$:%
\[
\chi_{ij}=-\frac{\partial\beta_{j}}{\partial\left\langle I^{i}\right\rangle }%
\]
is just the negative entropy Hessian $\hat{\chi}_{ij}=-\kappa_{ij}$
(\ref{eq1}) which obviously is not a second-rank covariant tensor. The
connection between the generalized response matrix $\chi^{ij}$ with the
entropy Hessian $\kappa_{ij}$ leads directly to rephrased the microcanonical
Maxwell identities (\ref{maxwell}) as the symmetric character of the
generalized response matrix:
\begin{equation}
\chi^{ij}\equiv\chi^{ji}.
\end{equation}

The ensemble equivalence within the generalized Boltzmann-Gibbs ensemble
(\ref{gce}) in the representation $\mathcal{R}_{I}$ demands the negative
definition of the entropy Hessian $\kappa_{ij}$, which is equivalent to demand
the positive definition of the generalized response matrix $\chi^{ij}$
(\ref{response matrix}). Thus, regions of ensemble equivalence are just
regions with anomalous behavior in the response functions $\chi^{ij}$.

The correlations matrixes $G^{ij}=\left\langle \delta I^{i}\delta
I^{j}\right\rangle $, $M_{i}^{j}=\left\langle \delta\beta_{i}\delta
I^{j}\right\rangle $ and $T_{ij}=\left\langle \delta\beta_{i}\delta\beta
_{j}\right\rangle $ are related by a set of thermodynamical identities which
could be referred as the \textit{generalized} \textit{fluctuation relations}:%
\begin{align}
\left\langle \delta\beta_{i}\delta I^{j}\right\rangle  &  =\delta_{i}%
^{i}+\kappa_{ij}\left\langle \delta I^{m}\delta I^{j}\right\rangle
,\label{ThF2}\\
\left\langle \delta I^{i}\delta I^{n}\right\rangle \left\langle \delta
\beta_{n}\delta\beta_{j}\right\rangle  &  =\left\langle \delta\beta_{j}\delta
I^{n}\right\rangle \left\langle \delta\beta_{n}\delta I^{i}\right\rangle ,
\label{ThF3}%
\end{align}
within the representation $\mathcal{R}_{I}$. The connection between the
entropy Hessian $\kappa_{ij}$ with the generalized response matrix $\chi^{ij}$
allows us to rephrase the first-generalized relation (\ref{ThF2}) as follows:
\begin{equation}
\chi^{ij}=\chi^{im}\left\langle \delta\beta_{m}\delta I^{j}\right\rangle
+\left\langle \delta I^{i}\delta I^{j}\right\rangle ,
\label{fluctuation-response}%
\end{equation}
which is the \textit{generalized response-fluctuation theorem}.

A fundamental difference of the above expressions in regard to the
conventional result (\ref{HG1}) is the presence of a non vanishing correlation
matrix $M_{i}^{j}=\left\langle \delta\beta_{i}\delta I^{j}\right\rangle $.
Moreover, the identity (\ref{ThF3}) has no counterpart within the conventional
Thermodynamics. Such correlations are not present within the ordinary
Boltzmann-Gibbs distribution (\ref{bgd}) since the canonical parameters of the
thermostat are keep fixed, so that $\delta\beta\equiv0\Rightarrow$
$\left\langle \delta\beta_{i}\delta I^{j}\right\rangle =\left\langle
\delta\beta_{n}\delta\beta_{j}\right\rangle \equiv0$. Such limitation does not
allow to the Boltzmann-Gibbs description to access to all those regions with a
anomalous behavior of the response functions, i.e. regions with a negative
heat capacity. As already shown by us in the refs.\cite{vel.geo,vel-mmc}, the
only way to access to such anomalous regions with ensemble inequivalence
within a Boltzmann-Gibbs-like description is by using a generalized thermostat
with a fluctuating effective canonical parameters $\tilde{\beta}=\eta
\cdot\partial\Theta/\partial I$ in order to control the thermodynamic
equilibrium of the interest system by keeping fixed the \textit{averages} of
the ordinary equilibrium conditions $\left\langle \tilde{\beta}_{i}%
\right\rangle =\left\langle \hat{\beta}_{i}\right\rangle .$

The generalized fluctuation relations (\ref{ThF2})\ admit another
interpretation. Let us consider a representation $\mathcal{R}_{J}$ where the
entropy Hessian be a diagonal matrix in a given point $J$, $\kappa_{ij}%
=\kappa_{i}\delta_{ij}$, which can be obtained from the representation
$\mathcal{R}_{I}$\ by using an appropriate orthogonal rotation. Whenever an
Hessian eigenvalue $\kappa_{i}>0$ (when the interest point belongs to a
regions with ensemble inequivalence) the corresponding correlation
$\left\langle \delta\beta_{i}\delta J^{i}\right\rangle =1+\kappa
_{i}\left\langle \left(  \delta J^{i}\right)  ^{2}\right\rangle $, and
consequently:
\begin{equation}
\left\langle \delta\beta_{i}\delta J^{i}\right\rangle \geq1. \label{unc}%
\end{equation}
Such inequality looks-like an \textit{uncertainly relation} which talks about
that the physical quantity $J^{i}$ and the corresponding canonical parameter
$\beta_{i}$ of the generalized thermostat can not be kept fixed when the
corresponding eigenvalue $\kappa_{i}$\ of the entropy Hessian be nonnegative,
that is, when the thermodynamical state of the interest system belongs to a
region of ensemble inequivalence. Thus, the quantities $\beta_{i}$ and $J^{j}$
can be considered within the present framework as \textit{complementary
thermodynamic quantities}. This interpretation allows us to refer the
generalized fluctuation identities (\ref{ThF2}) as \textit{uncertainly
thermodynamic relations} \cite{vel.geo}.

It is necessary before end this subsection to devote some comments about the
\textit{phase transitions}. The phase transitions within the conventional
Thermodynamics are mathematically identified by the non analyticities of the
thermodynamic potential which is relevant in a given application. Within the
microcanonical description of an isolated system the only relevant
thermodynamic potential is the microcanonical entropy $S$.

The microcanonical entropy $S$ is always a continuous function, and it is
analytical whenever the number of degrees of freedom of the interest system
$n$ be finite. Thus, as in the conventional Thermodynamics, the non
analyticities of the entropy per particle $s=S/n$ \textit{can appear only in
the thermodynamic limit} $n\rightarrow\infty$. Since the microcanonical
entropy is a scalar function from the viewpoint of the reparametrization
invariance, its nonanalyticities will appears in any representation
$\mathcal{R}_{\Theta}$ of the abstract space $\Im$, and therefore,
\textit{these mathematical anomalies are reparametrization invariant}. Thus,
the lost of analyticity of the microcanonical entropy is identified in the
present geometric framework as the mathematical signature of the
\textit{microcanonical phase transitions}. It is said microcanonical phase
transitions in order to distinguish them from other macroscopic anomalies
which are related with the occurrence of the phenomenon of ensemble
inequivalence in an open system. Generally speaking, there exist two generic
types of microcanonical phase transitions: the \textit{microcanonical
discontinuous phase transitions} (associated to a discontinuity in some of the
first derivatives of the microcanonical entropy $S$) and the
\textit{microcanonical continuous phase transitions} (associated to lost of
analyticity of the microcanonical entropy $S$ with continuous first derivatives).

The dynamical origin of the microcanonical ensemble leads to identify any
microcanonical phase transition as the occurrence of some kind of dynamical
anomaly in the microscopic picture of the system. The evidences analyzed in
the previous work \cite{vel.geo} suggest strongly that the microcanonical
phase transitions are directly related with the occurrence of the
\textit{ergodicity breaking phenomenon}, which takes place when the time
averages and the ensemble averages of certain macroscopic observables can not
be identified due to the microscopic dynamics is effectively trapped in
different subsets of the configurational or phase space during the imposition
of the thermodynamic limit $n\rightarrow\infty$ \cite{Gold}.

As elsewhere shown, the convexity of the microcanonical entropy leads to the
existence of a lost of analyticity of the Planck thermodynamical potential
$P\left(  \beta\right)  $ in the thermodynamic limit via the Legendre
transformation (\ref{LT1}). Such discontinuity in some of the first
derivatives of $P\left(  \beta\right)  $ is the signature of the first-order
phase transitions in the conventional Thermodynamics. It is necessary to
remark at this point that this kind of anomalies can not be microcanonically
relevant since they have only sense for the open systems instead of the
isolated ones \cite{vel.geo}. The first-order phase transitions are untimely
related with the phenomenon of ensemble inequivalence appearing as a
consequence of the non negative definition of the entropy Hessian $\kappa
_{ij}$ in a given representation $\mathcal{R}_{\Theta}$ of the abstract space
$\Im$. Therefore, the existence of the ensemble inequivalence depends
crucially on the nature of the external apparatus controlling the
thermodynamic equilibrium of the interest system. Thus, the first-order phase
transitions can be considered as \textit{avoidable thermodynamical anomalies
of the open systems}. This last feature can be used for enhancing the
available Monte Carlo methods based on the Statistical Mechanics in order to
avoid the ensemble inequivalence by considering the generalized
Boltzmann-Gibbs distributions (\ref{gce}).

I recommend to the interested reader to read\ the long discussion exposed in
the ref.\cite{vel.geo} for a better understanding about all these important
questions. I shall consider in the section \ref{apply} an example where the
existence of a first-order phase transition in the canonical description hides
the occurrence of two microcanonical phase transitions in the thermodynamical
description of the $q=10$ states Potts model system.

\subsection{Conjugate variables and the Riemannian interpretation of the
fluctuations}

The introduction of the dispersion $\delta\omega_{i}$:%
\begin{equation}
\delta\omega_{i}=\delta\beta_{i}-\kappa_{in}\delta I^{n},
\end{equation}
allows to rewrite the fundamental relation (\ref{gen1}) as follows:%
\begin{equation}
\left\langle \delta\omega_{i}\delta I^{n}\right\rangle =\delta_{i}^{n}.
\end{equation}
The correlations $\left\langle \delta\omega_{i}\delta\omega_{j}\right\rangle $
can be derived as follows: \
\begin{align}
&  =\left\langle \delta\beta_{i}\delta\omega_{j}\right\rangle -\kappa
_{in}\left\langle \delta I^{n}\delta\omega_{j}\right\rangle ,\nonumber\\
&  =\left\langle \delta\beta_{i}\delta\omega_{j}\right\rangle -\kappa
_{ij}\nonumber\\
&  =T_{ij}-\kappa_{jn}M_{i}^{n}-\kappa_{ij},
\end{align}
where the substitution of equation (\ref{a1}) and (\ref{r3}) leads finally to
the following result:%
\begin{equation}
\left\langle \delta\omega_{i}\delta\omega_{j}\right\rangle =G_{ij}.
\end{equation}

It is very easy to notice that the quantity $\omega_{i}$ is just the
difference between the effective canonical parameter $\tilde{\beta}=\eta
\cdot\partial\Theta/\partial I$ of the external apparatus $\wp_{\eta}$ and the
canonical parameter of the system $\hat{\beta}=\partial S/\partial I$,
$\omega_{i}\equiv\tilde{\beta}_{i}-\hat{\beta}_{i}$, that is:%
\begin{equation}
\omega_{i}=\eta_{k}\frac{\partial\Theta^{k}}{\partial I^{i}}-\frac{\partial
S}{\partial I^{i}}=\frac{\partial}{\partial I^{i}}P\left(  \Theta
\,;\eta\right)  ,
\end{equation}
where $P\left(  \Theta;\eta\right)  $ is the functional:%
\begin{equation}
P\left(  \Theta;\eta\right)  =\eta\cdot\Theta-S.
\end{equation}
The quantities $\omega=\left\{  \omega_{i}\right\}  $ and $I=\left\{
I^{i}\right\}  $ can be classified as \textit{conjugated variables }since they
obey the relations:
\begin{equation}
\left\langle \delta I^{i}\delta I^{j}\right\rangle =G^{ij},~\left\langle
\delta\omega_{i}\delta I^{n}\right\rangle =\delta_{i}^{n},~\left\langle
\delta\omega_{i}\delta\omega_{j}\right\rangle =G_{ij}. \label{conjugate}%
\end{equation}
In fact, the approximation (\ref{appb}) leads directly to the following
transformation rule between the fluctuations $\delta\omega$'s and $\delta
I$'s:
\begin{equation}
\delta\beta_{i}=F_{ij}\delta I^{i}\Rightarrow\delta\omega_{i}=G_{ij}\delta
I^{j},
\end{equation}
which allows us to obtain the second and third identities from the definition
of the correlation matrix $G^{ij}$ in the equations (\ref{conjugate}).

The generalized Boltzmann-Gibbs distribution (\ref{gce}) can be rephrased as follows:%

\begin{align}
d\omega_{GC}\left\{  I\right\}   &  =\frac{1}{Z\left(  \eta\right)  }%
\exp\left[  -\eta\cdot\Theta\left(  I\right)  +S\left(  I\right)  \right]
\frac{dI}{\delta I_{0}},\nonumber\\
&  \equiv\frac{1}{Z\left(  \eta\right)  \delta I_{0}}\exp\left[  -P\left(
\Theta;\eta\right)  \right]  dI.
\end{align}
The power expansion of the functional $P\left(  \Theta;\eta\right)  $ in the
neighborhood of its minimum is given by:
\begin{align}
P\left(  \Theta;\eta\right)   &  =P\left(  \Theta^{\ast};\eta\right)
+\frac{\partial P\left(  \Theta^{\ast};\eta\right)  }{\partial I^{i}}\delta
I^{i}+\nonumber\\
&  +\frac{1}{2}\frac{\partial^{2}P\left(  \Theta^{\ast};\eta\right)
}{\partial I^{i}\partial I^{j}}\delta I^{i}\delta I^{j}+\ldots\nonumber\\
&  =P^{\ast}+\omega_{i}^{\ast}\delta I^{i}+\frac{1}{2}G_{ij}\delta I^{i}\delta
I^{j}+\ldots
\end{align}
where%
\begin{equation}
\omega_{i}^{\ast}=\frac{\partial P\left(  \Theta^{\ast};\eta\right)
}{\partial I^{i}}=0,\text{~}G_{ij}=\frac{\partial^{2}P\left(  \Theta^{\ast
};\eta\right)  }{\partial I^{i}\partial I^{j}}. \label{minimum}%
\end{equation}
The Gaussian approximation allows us to estimate the canonical partition
function as follows:%
\begin{equation}
Z\left(  \eta\right)  \cong\exp\left[  -P^{\ast}\right]  \sqrt{\det\left(
2\pi G^{ij}\right)  }\delta I_{0}^{-1},
\end{equation}
which allows us to estimate finally the generalized Boltzmann-Gibbs
distribution in terms of the variable $I$ by:
\begin{equation}
d\omega_{GC}\left\{  I\right\}  =\frac{1}{\sqrt{\det\left(  2\pi
G^{ij}\right)  }}\exp\left\{  -\frac{1}{2}G_{ij}\delta I^{i}\delta
I^{j}\right\}  dI. \label{prob1}%
\end{equation}
The transformation from $I$ towards the conjugated variables $\omega$ allows
us to rewrite the above results as follows:%
\begin{equation}
d\omega_{GC}\left\{  \omega\right\}  =d\omega\int\delta\left\{  \omega
-\omega\left(  I\right)  \right\}  d\omega_{GC}\left\{  I\right\}  ,
\end{equation}
which leads immediately within the Gaussian approximation to the result:
\begin{equation}
d\omega_{GC}\left\{  \omega\right\}  \equiv\frac{1}{\sqrt{\det\left(  2\pi
G_{ij}\right)  }}\exp\left\{  -\frac{1}{2}G^{ij}\omega_{i}\omega_{j}\right\}
d\omega. \label{prob2}%
\end{equation}

The correlation tensor $G_{ij}$ or $G^{ij}$ arises as a consequence of the
interaction between the generalized thermostat ($F_{ij}$) and the interest
system ($\kappa_{ij}$) within the generalized Boltzmann-Gibbs description
($G_{ij}=F_{ij}-\kappa_{ij}$). Such tensor provides in this context a
Riemannian interpretation of the fluctuations, where the norm:%
\begin{equation}
\delta s^{2}=G_{ij}\delta I^{i}\delta I^{j}, \label{metric}%
\end{equation}
is directly related to the probability of occurrence of a given fluctuation
$\delta I=I-I^{\ast}$ around the thermodynamic equilibrium of the system in
accordance with the equation (\ref{prob1}). I refer to a Riemannian metric
since the correlation matrix $G_{ij}$ behaves as a second-rank tensor under a
given reparametrization $\upsilon:\mathcal{R}_{I}\rightarrow\mathcal{R}_{J}$
\textit{within the same representation }$\mathcal{R}_{\Theta}$\textit{ of the
generalized Boltzmann-Gibbs ensemble}. Taking into account $J=\upsilon\left(
I\right)  \Rightarrow$ $\Theta\left(  I\right)  =\Theta\left[  \upsilon
^{-1}\left(  J\right)  \right]  $, the expression of the correlation matrix in
the representation $\mathcal{R}_{J}~$:
\begin{align}
G_{\alpha\beta}  &  =\frac{\partial^{2}P\left(  \Theta^{\ast};\eta\right)
}{\partial J^{\alpha}\partial J^{\beta}}=\frac{\partial I^{i}}{\partial
J^{\alpha}}\frac{\partial I^{j}}{\partial J^{\beta}}G_{ij}+\frac{\partial
^{2}I^{i}}{\partial J^{\alpha}\partial J^{\beta}}\omega_{i}^{\ast}\nonumber\\
&  \equiv\frac{\partial I^{i}}{\partial J^{\alpha}}\frac{\partial I^{j}%
}{\partial J^{\beta}}G_{ij},
\end{align}
where the vanishing of the second terms takes place as a consequence of the
stationary conditions (\ref{minimum}). This result generalizes the other
Riemannian interpretations of the Thermodynamics obtained in the past
\cite{rupper}. While the covariant second-rank tensor $G_{ij}$ characterizes
the probabilities of the fluctuations of the physical quantities $I=\left\{
I^{i}\right\}  $ of the interest system, its corresponding contravariant
tensor $G^{ij}$ characterizes the probabilities of the fluctuations of the
conjugated variables $\omega=\left\{  \omega_{i}\right\}  $ in the
neighborhood of its expectation $\omega^{\ast}=\left\langle \omega
\right\rangle =\left\langle \tilde{\beta}\right\rangle -\left\langle
\hat{\beta}\right\rangle \equiv0$ in accordance with the equation
(\ref{prob2}), which provides us information about the unusual thermodynamic
equilibrium between the generalized thermostat and the interest system. Thus,
the conjugated character of the variables $\omega$ and $I$ is closely related
with the complementary character of the canonical parameters $\tilde{\beta}$
or $\hat{\beta}$ and their corresponding physical quantities of the interest
system $I$.

\section{A formal application\label{apply}}

Let us reconsider again the example about a ferromagnetic system under the
influence of an external magnetic field $B$ directed along the z-axis. Let us
to introduce a distinction between the internal Hamiltonian $\hat{H}$ (related
with the system internal energy $U$) and the total Hamiltonian $\hat{H}%
_{B}=\hat{H}-B\hat{M}_{z}$ (related with the total energy $E=U-BM_{z}$). As
already commented, the Boltzmann-Gibbs distribution of the system under these
conditions:%
\begin{equation}
\hat{\omega}_{G}=\frac{1}{Z\left(  \beta;B_{z}\right)  }\exp\left[  -\beta
\hat{H}_{B}\right]  , \label{v1}%
\end{equation}
can be reinterpreted as a Boltzmann-Gibbs distribution (BGE) with canonical
parameters $\beta$ and $\lambda=-\beta B$:
\begin{equation}
\hat{\omega}_{G}=\frac{1}{Z\left(  \beta;\lambda\right)  }\exp\left[
-\beta\hat{H}-\lambda\hat{M}_{z}\right]  . \label{v2}%
\end{equation}
The above observation allows to distinguish the usual microcanonical
description (IME):%
\begin{equation}
\hat{\omega}_{M}=\frac{1}{\Omega\left(  E;B\right)  }\delta\left(  E-\hat
{H}_{B}\right)  , \label{vm1}%
\end{equation}
from the \textit{detailed microcanonical description }(DME):
\begin{equation}
\hat{\omega}_{M}=\frac{1}{\Omega\left(  U,M_{z}\right)  }\delta\left(
U-\hat{H}\right)  \delta\left(  M_{z}-\hat{M}_{z}\right)  . \label{vm2}%
\end{equation}
While the canonical ensembles (\ref{v1}) and (\ref{v2}) are identical, the
microcanonical ensembles (\ref{vm1}) and (\ref{vm2}) are essentially
\ inequivalent. The above interpretation allows us to consider the modulus of
the external magnetic field $B$ or more exactly, the quantity $\lambda=-\beta
B$ as the canonical control parameter which is thermodynamically complementary
to the magnetization of the system $M_{z}$. Consequently, the macroscopic
description provided by ensemble (\ref{vm1}) can be considered as an
intermediate description between the DME (\ref{vm2}) and the BGE (\ref{v2}),
which leads to the following hierarchy among these ensembles: DME$\Rightarrow
$IME$\Rightarrow$BGE. This hierarchy presupposes that the description DME
contains more information than the IME description, and at the same time this
last one contains more information than the BGE description.

Let us consider the analysis of the above magnetic system in terms of the
thermodynamic formalism. The total differential of the microcanonical entropy
of the detailed ensemble (\ref{vm2}) $S\left(  U,M_{z}\right)  $ is simply
given by:%

\begin{equation}
dS=\beta dU+\lambda dM,
\end{equation}
where:%
\begin{equation}
\beta=\frac{\partial S}{\partial U},~\lambda=\frac{\partial S}{\partial M_{z}%
}\Rightarrow\left(  \frac{\partial\lambda}{\partial U}\right)  _{M_{z}%
}=\left(  \frac{\partial\beta}{\partial M_{z}}\right)  _{U}.
\end{equation}
The introduction of the total energy $U=E+BM_{z}$ allows to rewrite the above
expression as follows:%
\begin{equation}
dS=\beta\left(  dU-BdM\right)  =\beta\left(  dE+M_{z}dB\right)  ,
\end{equation}
from which are derived the relations:%
\begin{equation}
\bar{S}\left(  E;B\right)  \equiv S\left(  U,M_{z}\right)  , \label{wrong}%
\end{equation}%
\begin{equation}
\beta=\frac{\partial\bar{S}}{\partial E},~\beta M_{z}=\frac{\partial\bar{S}%
}{\partial B}\Rightarrow\left(  \frac{\partial\left(  \beta M_{z}\right)
}{\partial E}\right)  _{B}=\left(  \frac{\partial\beta}{\partial B}\right)
_{E}.
\end{equation}
Thus, the consideration of the total energy $E$ instead of the internal energy
$U$ corresponds to a reparametrization where $\left(  U,M_{z}\right)
\Rightarrow\left(  E,B\right)  $. Reader can notice that such transformation
does not correspond to the diffeomorphic transformation considered by the
reparametrization invariance of the microcanonical ensemble discussed in the
present work. Finally, the Planck potential $P\left(  \beta,\lambda\right)  $
of the BGE is derived from the DME microcanonical entropy $S\left(
U,M_{z}\right)  $ throughout the Legendre transformation as follows:%
\begin{equation}
P\left(  \beta,\lambda\right)  =\beta U+\lambda M_{z}-S\left(  U,M_{z}\right)
,
\end{equation}
which is equivalent to the one obtained from the IME microcanonical entropy
$\bar{S}\left(  E;B\right)  $ by considering the relations $\lambda=-\beta B$
and $E=U-BM_{z}$:%
\begin{align}
P\left(  \beta,\lambda\right)   &  =\beta U-\beta BM-S\left(  U,M_{z}\right)
\nonumber\\
&  =\beta E-\bar{S}\left(  E;B\right)  =P\left(  \beta;B\right)  .
\end{align}

The microcanonical partition functions $\Omega\left(  U,M_{z}\right)  $ and
$\Omega\left(  E;B\right)  $ are related as follows:
\begin{align}
\Omega\left(  E;B\right)   &  =\int\Omega\left(  U,M_{z}\right)  \delta\left(
U-E-BM_{z}\right)  dUdM_{z},\nonumber\\
&  =\int\Omega\left(  E+BM_{z},M_{z}\right)  dM_{z},
\end{align}
which leads to the following maximization problem in the thermodynamic limit:%
\begin{equation}
\bar{S}\left(  E;B\right)  =\sup_{M_{z}}\left\{  S\left(  E+BM_{z}%
,M_{z}\right)  \right\}  , \label{true}%
\end{equation}
whose stationary conditions are given by:%
\begin{align}
\frac{\partial S\left(  U,M\,_{z}\right)  }{\partial U}B+\frac{\partial
S\left(  U,M_{z}\right)  }{\partial M_{z}}  &  =0,\label{magnet}\\
B^{2}\frac{\partial^{2}S}{\partial U^{2}}+2B\frac{\partial^{2}S}{\partial
U\partial M_{z}}+\frac{\partial^{2}S}{\partial M_{z}^{2}}  &  <0.
\label{stability}%
\end{align}
The relation (\ref{true}) clarifies the precise meaning of the relation
(\ref{wrong}), which talks us that the reparametrization $\left(  U,M\right)
\rightarrow\left(  E,B\right)  $ in the thermodynamic limit is just a kind of
projection where it could be involved an \textit{important lost of
thermodynamic information}.

Taking into consideration that $\beta=\partial S/\partial U$ and
$\lambda=\partial S/\partial M_{z}$, the condition (\ref{magnet}) is just the
relation $\lambda=-\beta B$. On the other hand, the general solution of this
stationary condition leads to what could be called as the
\textit{microcanonical magnetization curve} at a given magnetic field $B$,
$M=M\left(  U;B\right)  $. Generally speaking, such microcanonical
magnetization curve could be a \textit{multivaluated} function of the internal
energy, that is, there could be more than one admissible value of the
magnetization $M_{z}$ at a given value of the internal energy $U$. However,
only those points of the microcanonical magnetization curve satisfying the
stability condition (\ref{stability}) will be observed within the IME
description. Such condition is related with the negative definition of the
entropy Hessian $\kappa_{ij}$, which can be rephrased by eliminating the
magnetic field $B$ as follows:
\begin{equation}
\xi=\lambda^{2}\kappa_{UU}+\lambda\beta\left(  \kappa_{UM}+\kappa_{MU}\right)
+\kappa_{MM}\beta^{2}<0, \label{chita}%
\end{equation}
where the stationary condition (\ref{magnet}) was taken into account. Thus,
all those points where $\xi>0$ can not be observed in the IME description, and
consequently, such possibility involves a lost of thermodynamic information in
regard to the one provided by DME description. Thus, some branches of the
microcanonical magnetization curve can disappear within the IME description,
leading in this way to the existence of discontinuities. Such discontinuities
can be considered as the signature of \textit{microcanonical discontinuous
phase transitions} within the IME description because of them represent
discontinuities in the first derivative%
\begin{equation}
\beta M_{z}=\frac{\partial\bar{S}}{\partial B}%
\end{equation}
of the IME microcanonical entropy $\bar{S}\left(  E;B\right)  $. The lost of
information is more significant in the BGE description, which demands now the
negative definition of the entropy Hessian:%
\begin{equation}
\kappa_{UU}<0,~\kappa_{MM}\kappa_{UU}-\kappa_{UM}\kappa_{MU}>0,
\label{negative}%
\end{equation}
which is more restrictive than the condition (\ref{chita}), that is, the
attainability of the condition (\ref{negative}) leads to the attainability of
(\ref{chita}), but the converse is not true.%

\begin{figure}
[t]
\begin{center}
\includegraphics[
height=3.218in,
width=3.5405in
]%
{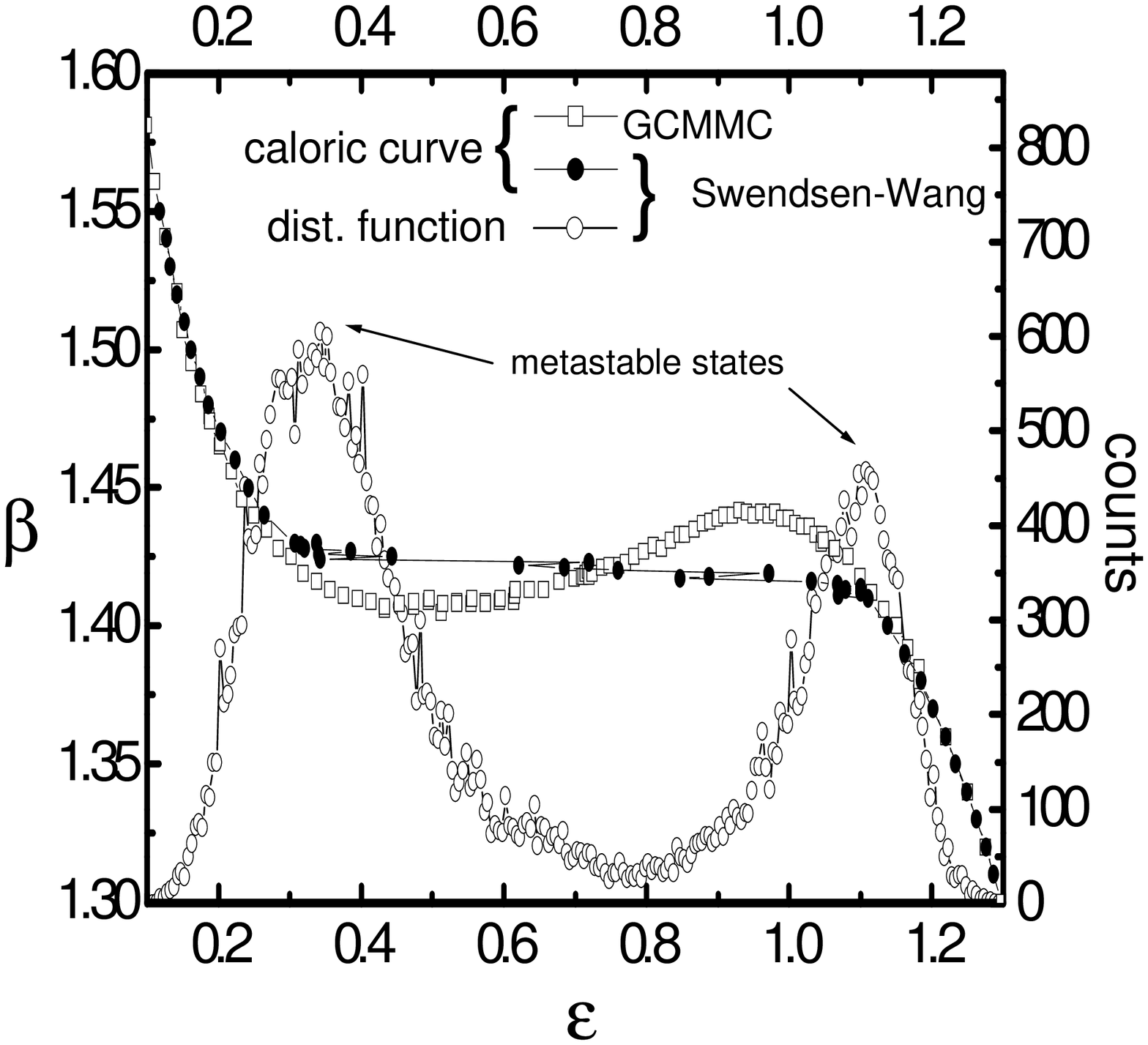}%
\caption{Comparative study between the Metropolis Monte Carlo based on the
generalized canonical ensemble (GCMMC)\ and the Swendsen-Wang (SW) cluster
algorithm during the thermodynamical description of the $q=10$ states Potts
model. The SW algorithm is unable to describe the microcanonical states with a
negative heat capacity during the first-order phase transitions since this
method is based on the consideration of the Gibbs canonical ensemble. This aim
is sucessfully performed by using the GCMMC algorithm, which is able to
explore all those energetic regions of ensemble inequivalence. Notice the
bimodal character of the energy distribution function within the Gibbs
canonical description at $\beta=1.42$, which is a feature of the first-order
phase transitions illustrating the competition among two metastable states
with different energies (coexistence phases). }%
\label{gc_sw.eps}%
\end{center}
\end{figure}
%

\begin{figure}
[t]
\begin{center}
\includegraphics[
height=3.1574in,
width=3.5405in
]%
{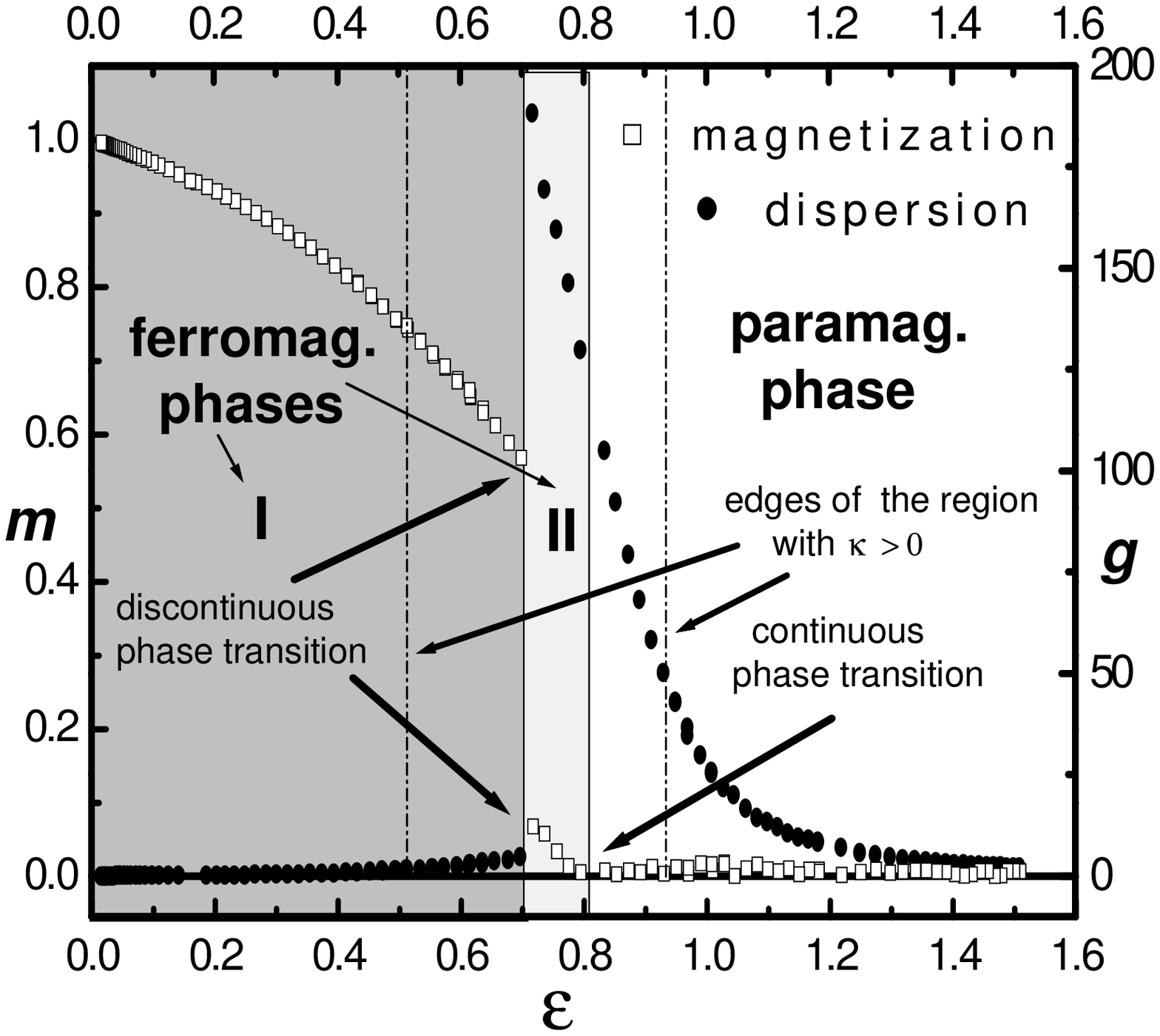}%
\caption{Magnetic properties of the $q=10$ states Potts model at zero magnetic
field $B$ within the IME microcanonical description in which is shown the
existence of two microcanonical phase transitions at $\varepsilon_{ff}%
\simeq0.7$ (ferro-ferro, discontinuous PT) and $\varepsilon_{fp}\simeq0.8$
(ferro-para, continuous PT). The dash-dot lines indicate the limit of the
energetic region with a negative heat capacity hidden by the ensemble
inequivalence.}%
\label{magnet1.eps}%
\end{center}
\end{figure}

The picture described above can be observed during the study of the
thermodynamic properties of the $q=10$ states Potts model \cite{pottsm}:%
\begin{equation}
\hat{H}=\sum_{\left\langle ij\right\rangle }\left(  1-\delta_{\sigma_{i}%
\sigma_{j}}\right)
\end{equation}
where the sum involves only neighbor-neighbor interactions on a square lattice
$L\times L$ with periodic boundary conditions. The spin variable of the
\textit{i-th} site $\sigma_{i}=1,2,...10$ are rephrased as bidimensional
vector $\mathbf{s}_{i}=\left[  \cos\left(  \omega\sigma_{i}\right)
,\sin\left(  \omega\sigma_{i}\right)  \right]  $ with $\omega=2\pi/q$, which
allows to introduce the microscopic magnetization as follows $\mathbf{M}%
=\sum_{i}\mathbf{s}_{i}$.

The study of this model system with $L=25$~by using the Gibbs canonical
ensemble (\ref{v1}) is characterized by the existence of a discontinuous phase
transition at $\beta_{c}\simeq1.42$ with $B=0$, which is recognized by the
existence of a plateau in the caloric curve $\beta$ \textit{versus}
$\varepsilon$ ($\varepsilon=E/N$ is the energy per particle with $N=L^{2}$)
from $\varepsilon_{1}=0.319$ to $\varepsilon_{2}=1.099$ corresponding to a
latent heat $q_{lt}=\varepsilon_{2}-\varepsilon_{1}\simeq0.78$. As already
commented, the first-order phase transition within the Gibbs canonical
ensemble is a consequence of an ensemble inequivalence between this
description and the IME description (\ref{vm1}), which takes place as a
consequence of the existence of thermodynamical states with a \textit{negative
specific heat}. These anomalous states manifest themselves as a backbending in
the caloric curve within the IME microcanonical ensemble (\ref{vm1}).

Reader can see all these behaviors in the FIG.\ref{gc_sw.eps}, which shows a
comparative study between two Monte Carlo simulations of this toy system at
$B=0$: the well-known Swedsen-Wang clusters algorithm (SW) based on the Gibbs
canonical ensemble \cite{wang2} and the Metropolis Monte Carlo algorithm
(GCMMC) based on the generalized Boltzmann-Gibbs ensemble (\ref{gce})
explained in the section \ref{open} and the ref.\cite{vel-mmc}. The SW
algorithm is unable to describe all those states with a negative heat
capacity, a task successfully performed by using the GCMMC algorithm. The key
for the success of this last Monte Carlo method is precisely the using of a
generalized Gibbs thermostat with a fluctuating inverse temperature
$\tilde{\beta}$. The caloric curve is obtained here by computing the average
values $\beta=\partial\bar{s}\left(  \varepsilon;B\right)  /\partial
\varepsilon=\left\langle \tilde{\beta}\right\rangle $ and $\varepsilon
=\left\langle \hat{\varepsilon}\right\rangle $, while the curvature or entropy
Hessian $\kappa=\partial^{2}\bar{s}\left(  \varepsilon;B\right)
/\partial\varepsilon^{2}$ can be derived from the unidimensional generalized
fluctuation relation:%
\begin{equation}
\left\langle \delta\beta\delta\varepsilon\right\rangle =1+\kappa\left\langle
\delta\varepsilon^{2}\right\rangle
\end{equation}
by computing the correlations $\left\langle \delta\beta\delta\varepsilon
\right\rangle $ and $\left\langle \delta\varepsilon^{2}\right\rangle $. Reader
can notice that no abrupt change of the energy appear by using the GCMMC
algorithm, which illustrates us the \textit{avoidable character of the
first-order phase transitions}, and therefore, the irrelevance of such
anomalies within a microcanonical description.

The study of the magnetic properties of this system within the IME
microcanonical ensemble (\ref{vm1}) at $B=0$ is shown in the
FIG.\ref{magnet1.eps}. This figure reveals the existence of two microcanonical
phase transitions which are hidden by the ensemble inequivalence within the
Gibbs canonical description. It is easy to show that such anomalies are
directly related with the lost of analyticity of the microcanonical entropy
$\bar{s}\left(  \varepsilon;B\right)  $\ in the thermodynamic limit as well as
the occurrence of ergodicity breaking in the microscopic picture of the
system. The interested reader can see more details about this study in the
refs.\cite{vel.geo,vel-mmc}. In spite of the significant amount of
thermodynamical information provided by the IME description (\ref{vm1}) in
regard to the BGE (\ref{v1}), the discontinuous character of the magnetization
curve is a clear indication about the lost of information of the IME
description in regard to the DME one (\ref{vm2}) appearing as a consequence of
the unattainability of the condition (\ref{chita}) for some macrostates. The
Monte Carlo simulations shows the existence of several metastable states with
different magnetizations in the energetic region $\left(  \varepsilon
_{A},\varepsilon_{B}\right)  $ with $\varepsilon_{A}\simeq0.7$ and
$\varepsilon_{B}\simeq0.93$, which indicates clearly the multivaluate
character of the microcanonical magnetization curve $M=M\left(  U;B=0\right)
$ in the DME description, which should exhibit an anomalous branch
characterized by an increasing of the magnetization with the increasing of the
internal energy $\partial M/\partial U>0$.

In complete analogy with the Gibbs canonical ensemble case, the existence of
metastable states provokes an exponential divergence of the correlation times
with the increasing of the system size during the Monte Carlo simulations, a
dynamical phenomenon referred\ in the literature as \textit{supercritical
slowing down}. The existence of this phenomenon is untimely related with the
lost of information associated to the ensemble inequivalence. To avoid this
difficulty is necessary to perform a thermodynamic description within the DME
ensemble. This aim could be carried out in an appropriated
\textit{canonical-way} within a generalized Boltzmann-Gibbs description
(\ref{gce}) whose generalized thermostat exhibits two fluctuating effective
control parameters $\tilde{\beta}$ and $\tilde{\lambda}$. This is just an
unusual equilibrium situation from the conventional Thermodynamics viewpoint
characterized by a fluctuating temperature $\tilde{T}=1/\tilde{\beta}$ and a
fluctuating external magnetic field $\tilde{B}=-\tilde{\lambda}/\tilde{\beta}$.

The dependence of the canonical parameters $\beta=\partial S/\partial U$ and
$\lambda=\partial S/\partial M_{z}$ with the microcanonical variables $\left(
U,M_{z}\right)  $ can be obtained by computing the averages
\begin{equation}
\beta=\left\langle \tilde{\beta}\right\rangle ,\lambda=\left\langle
\tilde{\lambda}\right\rangle ,U=\left\langle \hat{U}\right\rangle \text{ and
}M_{z}=\left\langle \hat{M}_{z}\right\rangle .
\end{equation}
The components of the entropy Hessian $\kappa_{ij}=\left\{  \kappa_{UU}%
,\kappa_{UM},\kappa_{MU},\kappa_{MM}\right\}  $ can be derived from the
generalized fluctuations relations:%
\begin{align}
\left\langle \delta\beta\delta U\right\rangle  &  =1+\kappa_{UU}\left\langle
\delta U^{2}\right\rangle +\kappa_{UM}\left\langle \delta M_{z}\delta
U\right\rangle ,\\
\left\langle \delta\beta\delta M_{z}\right\rangle  &  =\kappa_{UU}\left\langle
\delta U\delta M_{z}\right\rangle +\kappa_{UM}\left\langle \delta M_{z}%
^{2}\right\rangle ,\\
\left\langle \delta\lambda\delta U\right\rangle  &  =\kappa_{MU}\left\langle
\delta U^{2}\right\rangle +\kappa_{MM}\left\langle \delta M_{z}\delta
U\right\rangle ,\\
\left\langle \delta\lambda\delta M_{z}\right\rangle  &  =1+\kappa
_{MU}\left\langle \delta U\delta M_{z}\right\rangle +\kappa_{MM}\left\langle
\delta M_{z}^{2}\right\rangle ,
\end{align}
by computing the four components of the correlation matrix $M_{i}^{j}=\left\{
\left\langle \delta\beta\delta U\right\rangle ,\left\langle \delta\beta\delta
M_{z}\right\rangle ,\left\langle \delta\lambda\delta U\right\rangle
,\left\langle \delta\lambda\delta M_{z}\right\rangle \right\}  $ and the four
component of the correlation matrix $G^{ij}=\left\{  \left\langle \delta
U^{2}\right\rangle ,\left\langle \delta U\delta M_{z}\right\rangle
,\left\langle \delta M_{z}\delta U\right\rangle ,\left\langle \delta M_{z}%
^{2}\right\rangle \right\}  $. The microcanonical entropy $S\left(
U,M_{z}\right)  $ can be derived from the direct numerical integration of the
canonical parameters $\left(  \beta,\lambda\right)  $ and the entropy Hessian
$\kappa_{ij}$. The possibility of perform such study will be accounted in a
forthcoming paper.

As already shown in the subsection \ref{formalism}, the above generalized
fluctuation relations can be rewritten by introducing the generalized response
matrix $\chi^{ij}$:%

\begin{align}
\chi^{UU}  &  =-\frac{\partial U}{\partial\beta},~\chi^{UM}=-\frac{\partial
U}{\partial\lambda},\\
\chi^{MU}  &  =-\frac{\partial M_{z}}{\partial\beta},~\chi^{MM}=-\frac
{\partial M_{z}}{\partial\lambda},
\end{align}
as follows:
\begin{align}
\chi^{UU}  &  =\left\langle \delta U^{2}\right\rangle +\chi^{UU}\left\langle
\delta\beta\delta U\right\rangle +\chi^{UM}\left\langle \delta\lambda\delta
U\right\rangle ,\\
\chi^{UM}  &  =\left\langle \delta U\delta M\right\rangle +\chi^{UU}%
\left\langle \delta\beta\delta M\right\rangle +\chi^{UM}\left\langle
\delta\lambda\delta M\right\rangle ,\\
\chi^{MU}  &  =\left\langle \delta M\delta U\right\rangle +\chi^{MU}%
\left\langle \delta\beta\delta U\right\rangle +\chi^{MM}\left\langle
\delta\lambda\delta U\right\rangle ,\\
\chi^{MM}  &  =\left\langle \delta M^{2}\right\rangle +\chi^{MU}\left\langle
\delta\beta\delta M\right\rangle +\chi^{MM}\left\langle \delta\lambda\delta
M\right\rangle .
\end{align}
The above expressions exhibit the intrinsic reparametrization invariance of
the geometrical theory developed in the present work. However, conventional
Thermodynamics usually deals with the total energy $E=U-BM_{z}$ and the
external magnetic field $B$ instead of the internal energy $U$ and the
magnetization $M_{z}$. This is the reason why it is very interesting to
rephrased the above results in terms of $E$ and $B$.

Let us began with the components of the generalized response matrix, which can
be rephrased as:%

\begin{align}
\chi^{UU}  &  =-\frac{\partial E}{\partial\beta}+B\frac{\partial M_{z}%
}{\partial\beta}+\frac{1}{\beta}\left(  \frac{\partial E}{\partial B}%
+B\frac{\partial M_{z}}{\partial B}+M_{z}\right)  ,\\
\chi^{UM}  &  =\frac{1}{\beta}\left(  \frac{\partial E}{\partial B}%
+B\frac{\partial M_{z}}{\partial B}+M_{z}\right)  ,\\
\chi^{MU}  &  =-\frac{\partial M_{z}}{\partial\beta}+\frac{1}{\beta}%
B\frac{\partial M_{z}}{\partial B},~\chi^{MM}=\frac{1}{\beta}\frac{\partial
M_{z}}{\partial B},
\end{align}
where the symmetric property $\chi^{UM}=\chi^{MU}$ \ leads to the well-known
thermodynamical identity:
\begin{equation}
\frac{\partial E}{\partial B}=-\beta\frac{\partial M_{z}}{\partial\beta}%
-M_{z}\equiv T\frac{\partial M_{z}}{\partial T}-M_{z}.
\end{equation}
The fluctuations $\delta U$ should be rewritten in terms of the fluctuations
$\delta E$, where $\delta U=\delta E+B\delta M_{z}+M_{z}\delta B$. The
ordinary response-fluctuation relations corresponding to $\delta\beta=\delta
B=0$ are given by:%
\begin{align}
\tilde{\chi}^{EE}  &  =-\frac{\partial E}{\partial\beta}=\left\langle \delta
E^{2}\right\rangle ,~\tilde{\chi}^{MM}=\frac{1}{\beta}\frac{\partial M_{z}%
}{\partial B}=\left\langle \delta M_{z}^{2}\right\rangle ,\\
\tilde{\chi}^{EM}  &  =\frac{1}{\beta}\left(  \frac{\partial E}{\partial
B}+M_{z}\right)  =\left\langle \delta E\delta M_{z}\right\rangle =\tilde{\chi
}^{ME}=-\frac{\partial M_{z}}{\partial\beta},
\end{align}
where it was introduced what can be considered as the components of the
response matrix $\tilde{\chi}^{ij}=\left\{  \tilde{\chi}^{EE},\tilde{\chi
}^{EM},\tilde{\chi}^{ME},\tilde{\chi}^{MM}\right\}  $ in the representation
$\left(  E,M_{z}\right)  $.\ The reader can notice that $\tilde{\chi}%
^{EE}=T^{2}C$ and $\tilde{\chi}^{MM}=T\chi_{B}$, where $C$ and $\chi_{B}$ are
the heat capacity and the magnetic susceptibility respectively. These
classical expressions lost their validity when the response functions present
some anomalous behavior, like $C<0$ or $\chi_{B}<0$. As already commented,
this problem could be solved within the fluctuation theory by considering the
generalized response-fluctuation relations when $\delta\beta\not =0$ and
$\delta B\not =0$, which can be expressed after some algebra as follows:%
\begin{align}
\tilde{\chi}^{EE}  &  =\left\langle \delta Q^{2}\right\rangle +\tilde{\chi
}^{EE}\left\langle \delta\beta\delta Q\right\rangle +\tilde{\chi}%
^{EM}\left\langle \delta\kappa\delta Q\right\rangle ,\label{GEM1}\\
\tilde{\chi}^{EM}  &  =\left\langle \delta Q\delta M_{z}\right\rangle
+\tilde{\chi}^{EE}\left\langle \delta\beta\delta M_{z}\right\rangle
+\tilde{\chi}^{EM}\left\langle \delta\kappa\delta M_{z}\right\rangle ,\\
\tilde{\chi}^{ME}  &  =\left\langle \delta M_{z}\delta Q\right\rangle
+\tilde{\chi}^{ME}\left\langle \delta\beta\delta Q\right\rangle +\tilde{\chi
}^{MM}\left\langle \delta\kappa\delta Q\right\rangle ,\\
\tilde{\chi}^{MM}  &  =\left\langle \delta M_{z}^{2}\right\rangle +\tilde
{\chi}^{ME}\left\langle \delta\beta\delta M_{z}\right\rangle +\tilde{\chi
}^{MM}\left\langle \delta\kappa\delta M_{z}\right\rangle , \label{GEM2}%
\end{align}
being $\delta Q=\delta E+M_{z}\delta B$ and $\delta\kappa=-\beta\delta B$,
where $\delta Q$ characterizes the fluctuations of the transferred heat among
the interest system and the generalized thermostat.

The equations (\ref{GEM1}-\ref{GEM2}) constitutes the generalized
response-fluctuation relations in terms of the usual control parameters of the
conventional Thermodynamics for a magnetic system, $\left(  E,M_{z}%
;\beta,B\right)  $. The corresponding generalized expressions for a fluid
system is very easy to obtain by considering the correspondence $\left(
E,M_{z};\beta,B\right)  \leftrightarrow\left(  E,p;\beta,V\right)  $ where $p$
is the pressure and $V$ the volume of the container:%
\begin{align}
\tilde{\chi}^{EE}  &  =-\frac{\partial E}{\partial\beta},~\tilde{\chi}%
^{pp}=\frac{1}{\beta}\frac{\partial p}{\partial V},\\
\tilde{\chi}^{Ep}  &  =\frac{1}{\beta}\left(  \frac{\partial E}{\partial
V}+p\right)  ,~\tilde{\chi}^{pE}=-\frac{\partial p}{\partial\beta},
\end{align}
where the symmetric condition $\tilde{\chi}^{Ep}=\tilde{\chi}^{pE}$ leads to
the\ well-known thermodynamic identity:
\begin{equation}
\frac{\partial E}{\partial V}=-\beta\frac{\partial p}{\partial\beta}-p\equiv
T\frac{\partial p}{\partial T}-p.
\end{equation}
The generalized response-fluctuation relations are given now by:%
\begin{align}
\tilde{\chi}^{EE}  &  =\left\langle \delta Q^{2}\right\rangle +\tilde{\chi
}^{EE}\left\langle \delta\beta\delta Q\right\rangle +\tilde{\chi}%
^{Ep}\left\langle \delta\kappa\delta Q\right\rangle ,\\
\tilde{\chi}^{Ep}  &  =\left\langle \delta Q\delta p\right\rangle +\tilde
{\chi}^{EE}\left\langle \delta\beta\delta p\right\rangle +\tilde{\chi}%
^{Ep}\left\langle \delta\kappa\delta p\right\rangle ,\\
\tilde{\chi}^{pE}  &  =\left\langle \delta p\delta Q\right\rangle +\tilde
{\chi}^{pE}\left\langle \delta\beta\delta Q\right\rangle +\tilde{\chi}%
^{pp}\left\langle \delta\kappa\delta Q\right\rangle ,\\
\tilde{\chi}^{pp}  &  =\left\langle \delta p^{2}\right\rangle +\tilde{\chi
}^{pE}\left\langle \delta\beta\delta p\right\rangle +\tilde{\chi}%
^{pp}\left\langle \delta\kappa\delta p\right\rangle ,
\end{align}
where $\delta Q=\delta E+p\delta V$ and $\delta\kappa=-\beta\delta V$.

\section{Summary}

It has been considered a generalization of the Boltzmann-Gibbs distributions
(\ref{bgd}) based on the \ reparametrization invariance of the microcanonical
ensemble. The resulting distribution functions correspond to an equilibrium
situation where an interest system is put in contact with a
\textit{generalized thermostat} in order to keep fixed the average values of
the quantities $\hat{\Theta}=\Theta\left(  \hat{I}\right)  $: an external
control apparatus becoming equivalent for the case of a large enough interest
system to the ordinary thermostat with \textit{effective fluctuating canonical
parameters} $\tilde{\beta}=\eta\cdot\partial\Theta/\partial I$. It is shown
that the ordinary equilibrium condition between the canonical parameters of
the thermostat $\tilde{\beta}$ and the corresponding canonical parameters of
interest system $\hat{\beta}=\partial S/\partial I$ is only satisfied now in
average $\left\langle \tilde{\beta}\right\rangle =\left\langle \hat{\beta
}\right\rangle $ and these ideas are used for enhance the possibilities of the
well-known Metropolis importance sample algorithm in regions characterized by
the ensemble inequivalence.

The generalized Boltzmann-Gibbs ensemble leads in a natural way towards a
suitable extension of the classical fluctuation theory of the conventional
Thermodynamics by using a non Riemannian geometric framework which accounts
for the reparametrization changes within the microcanonical description.
Surprisingly, the present approach\ leads to a novel interpretation of the
thermodynamic states with negative specific heat and other anomalous behavior
of the response functions as macrostates which can be only controlled by an
external apparatus \textit{with fluctuating control parameters} $\tilde{\beta
}$. Thus, the generalized fluctuation relations (\ref{ThF2})\ act as certain
kind of thermodynamic uncertainly relations where the physical observables of
the interest system $I$ and the corresponding effective canonical parameters
of the external generalized thermostat $\tilde{\beta}$ behave as
\textit{complementary thermodynamical quantities}. The above results
constitute the basis what could be considered a generalized thermodynamic
formalism within the microcanonical description. A rephrasing the generalized
fluctuation relations is carried out with the introduction of the concept of
the conjugate canonical variables $\left(  \omega,I\right)  $ where
$\omega=\tilde{\beta}-\hat{\beta}$, a representation which possibilities a
Riemannian interpretation of the fluctuations within the generalized
Boltzmann-Gibbs description (\ref{gce}), constituting in this way a natural
extension of other geometric interpretations of the classical fluctuation
theory developed in the past \cite{rupper}. As example of application, it was
obtained the generalized response-fluctuation relations for a magnetic system
(and a fluid system) in terms of the natural control parameters of the
geometric framework, as well as the usual control parameters of the
conventional Thermodynamics.

\end{document}